\documentclass[fleqn,usenatbib]{mnras}

\usepackage{newtxtext,newtxmath}
\usepackage{mathrsfs}

\usepackage[T1]{fontenc}
\usepackage{ae,aecompl}

\usepackage{graphicx}	
\usepackage{amsmath}	
\usepackage{amssymb}	

\title[A young galaxy cluster in the old Universe]{A young galaxy cluster in the old Universe}

\author[T. Hashimoto et al.]{
Tetsuya Hashimoto,$^{1}$\thanks{E-mail: tetsuya@phys.nthu.edu.tw}
Tomotsugu Goto,$^{1}$
Rieko Momose,$^{2}$
Chien-Chang Ho,$^{1}$
\newauthor
Ryu Makiya,$^{3}$
Chia-Ying Chiang$^{1}$
and Seong Jin Kim$^{1}$
\\
$^{1}$Institute of Astronomy, National Tsing Hua University, 101, Section 2. Kuang-Fu Road, Hsinchu, 30013, Taiwan (R.O.C.)\\
$^{2}$Department of Astronomy, School of Science, The University of Tokyo, 7-3-1 Hongo, Bunkyo-ku, Tokyo, 113-0033, Japan\\
$^{3}$Kavli Institute for the Physics and Mathematics of the Universe (Kavli IPMU, WPI), Todai Institutes for Advanced Study, \\The University of Tokyo, Kashiwa 277-8583, Japan\\
}
\date{Accepted 2019 August 2. Received 2019 August 2; in original form 2019 March 21}

\pubyear{2019}

\begin{document}
\label{firstpage}
\pagerange{\pageref{firstpage}--\pageref{lastpage}}
\maketitle

\begin{abstract}
Galaxies evolve from a blue star-forming phase into a red quiescent one by quenching their star formation activity.
In high density environments, this galaxy evolution proceeds earlier and more efficiently.
Therefore, local galaxy clusters are dominated by well-evolved red, elliptical galaxies.
The fraction of blue galaxies in clusters monotonically declines with decreasing redshift, i.e., the Butcher-Oemler effect.
In the local Universe, observed blue fractions of massive clusters are as small as $\lesssim$ 0.2.
Here we report a discovery of a \lq \lq blue cluster\rq \rq, that is a local galaxy cluster with an unprecedentedly high fraction of blue star-forming galaxies yet hosted by a massive dark matter halo.
The blue fraction is 0.57, which is 4.0 $\sigma$ higher than those of the other comparison clusters under the same selection and identification criteria.
The velocity dispersion of the member galaxies is 510 km s$^{-1}$, which corresponds to a dark matter halo mass of 2.0$^{+1.9}_{-1.0}\times 10^{14}$ M$_{\odot}$.
The blue fraction of the cluster is more than 4.7 $\sigma$ beyond the standard theoretical predictions including semi-analytic models of galaxy formation.
The probability to find such a high blue fraction in an individual cluster is only 0.003\%, which challenges the current standard frameworks of the galaxy formation and evolution in the $\Lambda$CDM Universe.
The spatial distribution of galaxies around the blue cluster suggests that filamentary cold gas streams can exist in massive halos even in the local Universe.
However these cold streams have already disappeared in the theoretically simulated local universes. 
\end{abstract}

\begin{keywords}
galaxies: clusters: individual:SDSS-C4 3028  -- galaxies: evolution -- galaxies: formation -- galaxies: star formation
\end{keywords}


\section{Introduction}
\label{introduction}
Galaxies evolve from a blue star-forming phase into a red quiescent one by quenching their star formation activity. 
In fact, the number fraction of blue star-forming galaxies in clusters declines with decreasing redshift \citep[e.g.,][]{Butcher1984}.
Environmental effects on galaxy evolution are more efficient in galaxy clusters than the field, through physical processes such as virial shocks, strangulation, and gas stripping \citep[e.g.,][]{Birnboim2003,Peng2015,Gunn1972}.
Therefore, the most evolved galaxy systems can be found in cluster environments in the local Universe.
The typical observed blue fraction of massive clusters is only $\lesssim$ 0.2 in the local Universe \citep[e.g.,][]{Butcher1984}.

So far, great efforts have been made to identify higher redshift galaxy clusters, targeting not only red galaxies but also star-forming emission-line galaxies \citep[e.g.,][]{Koyama2010,Hayashi2012,Hayashi2018,Oguri2018}.
In contrast to the local Universe, high-$z$ clusters obviously include a number of star-forming galaxies \citep[e.g.,][]{Hayashi2018}.
A possible explanation for the high star-forming fractions in high-$z$ clusters is cold streams accreting into cluster halo.
Cosmological hydrodynamical simulations of galaxy formation suggest that the cold streams can contribute to the intense star formation in cluster environments in the high-$z$ Universe \citep[e.g.,][]{Dekel2009}.
In order to trigger star formation, the cold streams need to overcome the virial shocks that heat up accreting gas. 
The virial shocks are invoked by the huge gravitational potential of a hosting halo and creates hot gas in the intercluster medium \citep[e.g.,][]{Birnboim2003}.
The balance between cold streams and the virial shocks roughly determines the star formation activity in clusters from the point of view of simulations.
Massive clusters, e.g., $\sim 10^{14}$ M$_{\odot}$, can host cold streams only in the high-$z$ Universe \citep[e.g.,][]{Dekel2006,Dekel2009}.
The cold streams do not survive in massive halos in the local Universe, because the virial shocks are relatively more efficient \citep{Dekel2009}.
Therefore local massive galaxy clusters contain few star-forming galaxies in general.

Apart from the general understanding of galaxy populations in local cluster environments, some galaxy clusters show high fractions of star-forming galaxies. 
\citet{Oemler1974} investigated number fractions of spiral galaxies in 15 local rich clusters. 
The reported fractions are between 0.18 and 0.63 with a median value of 0.28. 
\citet{Propris2004} investigated blue fractions of 60 local clusters identified in the Two-Degree Field Galaxy Redshift Survey \citep[2dFGRS;][]{Colless2001}. 
They defined blue galaxies based on colour magnitude diagrams. 
The blue fractions distribute from 0.0 to 0.6 with a mean value of 0.13. 
\citet{Campusano2018} updated 2dFGRS cluster identification and adopted a late-type/early-type classification by a parameter, $\eta$. 
This parameter is derived from spectroscopic features, e.g., emission and absorption lines \citep{Madgwick2003}. 
\citet{Campusano2018} concluded that among 207 identified clusters at $z=0.04-0.09$, 63\% are dominated by early-type galaxies and 37\% are dominated by late-type galaxies. 
Recently \citet{Ai2018} investigated spiral fractions of the Sloan Digital Sky Survey Data Release 7 \citep[SDSS DR7;][]{Abazajian2009} local clusters identified by \citet{Berlind2006}. 
They used a visual classification by the Galaxy Zoo 1 \citep{Lintott2011} to identify spiral galaxies. 
The spiral fractions show a large dispersion from 0.1 to 0.9 with a mean value of ~0.5.

These previous studies are based on different definitions of \lq\lq blue galaxies\rq \rq\ and different galaxy samples with different magnitude limits. 
All the differences between these analyses affect the blue fraction. 
Therefore, direct comparison of blue fractions between different approaches and also comparison with galaxy evolution models are not straightforward. 
In addition, the previous studies include as many clusters as possible, resulting in a mixture of diverse cluster properties controlling the blue fraction such as density environment, halo mass, and dynamical stage \citep[e.g.,][]{Butcher1984,Ai2018}.
The variety of definitions of the blue fraction and the mixture of diverse cluster properties have prevented us from correctly understanding how unusual blue clusters are in the local Universe.

One simple way to highlight the existence of local blue clusters is to focus on the most evolved systems in the Universe, i.e., local clusters in very high dense environments hosted by massive and dynamically relaxed dark matter halos. 

In this paper we report the discovery of a local \lq \lq blue cluster\rq \rq\ that is hosted by a massive and dynamically relaxed halo.
The density environment is $\sim 8 \sigma$ higher than the median value.
Here we use the \lq \lq main sequence\rq \rq\ to define blue star-forming galaxies, i.e, stellar mass-star formation rate (SFR) plane. 
The main sequence is a more quantitative parameter to describe star formation activity than morphological and colour classifications in previous studies. 
It is also predictable in a majority of galaxy evolution models, which allowed us to directly compare observations and models with exactly same criteria and analysis. 
The blue cluster consists of an unprecedentedly high fraction of star-forming galaxies, in contrast to the current theoretical predictions.

The structure of this paper is as follows.
In Section \ref{sample_selection} we describe our observational sample selection criteria.
Data analysis including cluster identification and our theoretical comparison sample are demonstrated in Section \ref{analysis}.
In Section \ref{results} we show physical properties of the blue cluster including the blue fraction that is located more than 4.0 $\sigma$ beyond the comparison sample.
Possible physical explanations are discussed in Section \ref{discussion} followed by conclusions in Section \ref{conclusion}.
Throughout this paper, we assume a cosmology of ($\Omega_{m}$,$\Omega_{\Lambda}$,$\Omega_{b}$,$h$)=(0.307, 0.693, 0.0486, 0.677) by {\it Planck15} \citep{Planck2015}, unless otherwise mentioned.

\section{Sample selection}
\label{sample_selection}
\begin{figure}
	\includegraphics[width=\columnwidth]{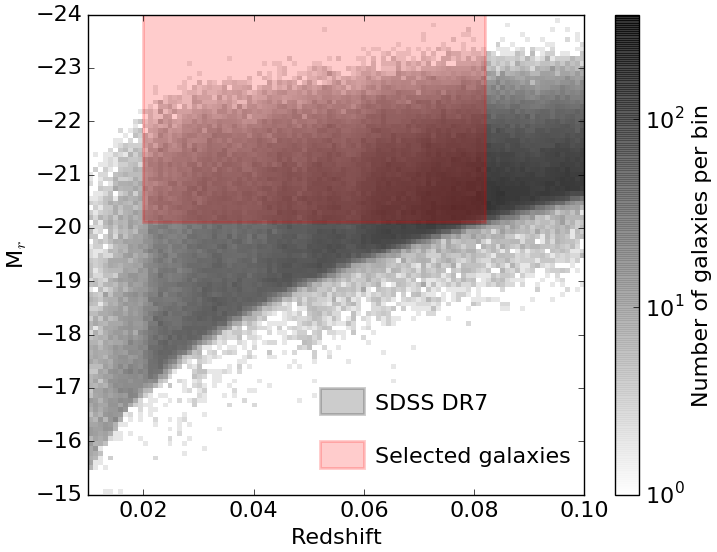}
    \caption{The selection box for our sample in the SDSS DR7.
    The grey scale shows the number density of galaxies.
    Galaxies inside the red box are selected as our volume-limited sample.
    }
    \label{newfig1}
\end{figure}

We constructed a volume-limited galaxy sample selected from the SDSS DR7 \citep{Abazajian2009} with redshift between 0.02 and 0.082 and $r$-band absolute magnitude, M$_{r}$, brighter than M$_{r}^{*}$+1.5=-20.1 mag.
Here M$_{r}^{*}$=-21.6 mag \citep{Blanton2001} is the magnitude at the knee of the galaxy luminosity function.
The r-band absolute magnitude, M$_{r}$, is calculated by following 
\begin{equation}
M_{r}=m_{r}-5.0(\log_{10}D_{L}-1.0)-K_{\rm Corr},
\end{equation}
where $m_{r}$, $D_{L}$ and $K_{Corr}$ are the apparent r-band magnitude, luminosity distance and $K$-correction \citep{Chilingarian2012}, respectively.
The selection box is designed to avoid the incompleteness of the SDSS spectroscopic observations both in the faint galaxies at higher redshift and in the brighter galaxies at lower redshift (Fig. \ref{newfig1}). 

To separate galaxies into two populations, blue star-forming and red quiescent, we used the so-called main sequence of galaxies in the stellar mass-SFR plane.
We used the publicly available catalogue\footnote[1]{https://wwwmpa.mpa-garching.mpg.de/SDSS/} of SFR and stellar mass produced in collaboration between the Max Planck Institute for Astrophysics and Johns Hopkins University \citep{Kauffmann2003,Brinchmann2004,Tremonti2004,Salim2007} based on the SDSS DR7. 
Among the galaxies in the catalogue, we selected objects with the type of GALAXY, and removed duplications and objects with any of suspicious flags for photometry, redshift, and stellar mass.
The sample contains 142,856 galaxies in total.\\

\section{Analysis}
\label{analysis}
\subsection{Identification of galaxy clusters}
\label{identification}
\begin{figure}
	\includegraphics[width=\columnwidth]{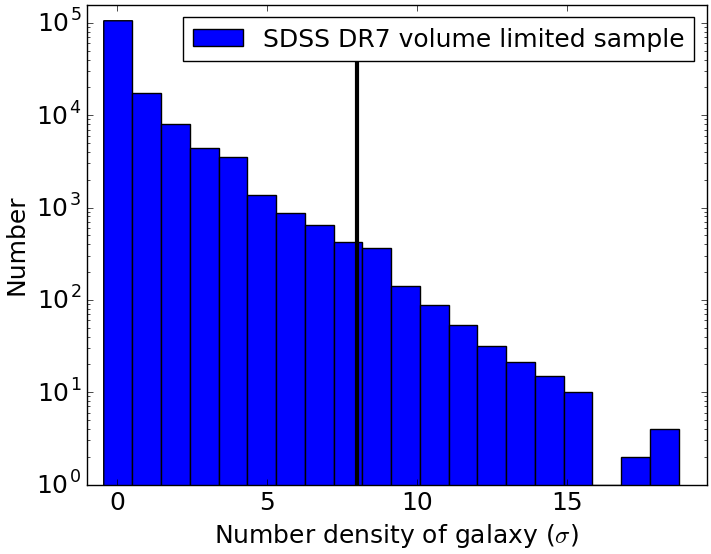}
    \caption{
    A histogram of the galaxy number density significance.
    The vertical line corresponds to 8 $\sigma$ from the median of galaxy number density.
    The galaxies beyond the vertical line are selected as starting points of the friends-of-friends method to identify galaxy clusters.
    }
    \label{newfig2}
\end{figure}

In our sample we assigned a projected number density to the individual galaxies as $\delta={\rm N}/(\pi r_{\rm link}^{2}$), where N is the number of galaxies within a projected linking length ($r_{\rm link}$) of 0.75 Mpc, and relative velocity $\pm$ 1000 km s$^{-1}$.
The histogram of the number density of our sample is shown in Fig. \ref{newfig2}.
We selected galaxies in the densest environments, which are higher than 8$\sigma$ from the median of $\delta$ (vertical solid line in Fig. \ref{newfig2}).
A simple friends-of-friends method was applied to identify clusters in the very dense environments.
In the friends-of-friends method, we adopted a linking length of 0.75 Mpc and relative velocity of $\pm$1000 km s$^{-1}$, because 0.75 Mpc is the typical size of galaxy clusters with $\sim10^{12}-10^{14}$ M$_{\odot}$.
For example, \citet{Tempel2014} reported that a majority of clusters with $\sim$10$^{12}$-10$^{14}$ M$_{\odot}$ have radii up to $\sim$0.75 Mpc.
The histogram of distances between each galaxy and its neighbours within $\pm$1000 km s$^{-1}$ peaks at 0.97 Mpc for our sample.
This peak distance is a representative projected distance between field galaxies.
Therefore, our choice of 0.75 Mpc is smaller than the typical distance between field galaxies. 
The linking length of 0.75 Mpc is a conservative length that gives a consistent cluster identification between flux-limited and volume-limited (M$_{r} <-20.1$) samples \citep{Tago2010}.

\begin{figure}
	\includegraphics[width=\columnwidth]{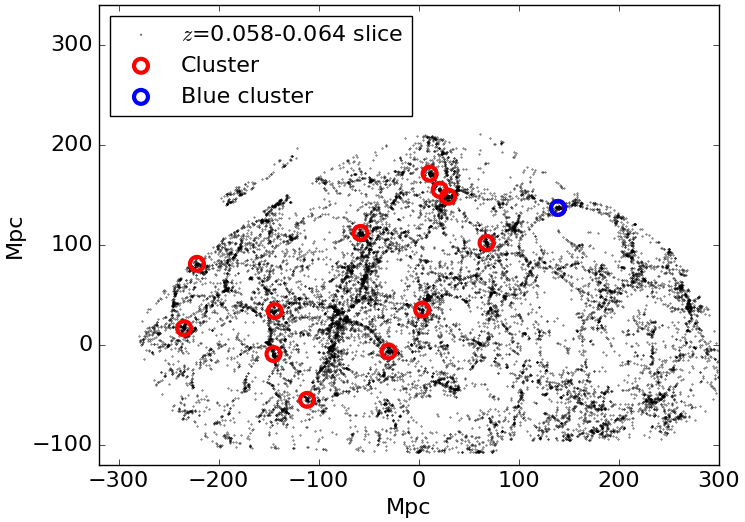}
    \caption{
    The spatial distribution of clusters in a redsfhit slice.
    The central positions of identified galaxy clusters are marked by red circles.
    The blue cluster is shown by a blue circle.
    }
    \label{newfig3}
\end{figure}

When a galaxy has a neighbour that satisfies these criteria, neighbours of the neighbour are searched using the same linking length and $\pm$1000 km s$^{-1}$ until no further neighbours are found.
The spacial distribution of the identified clusters are demonstrated in Fig. \ref{newfig3} in a redshift slice between $z=0.058$ and 0.064 that includes the blue cluster we found.
In Fig. \ref{newfig3}, the cluster position and redshift are determined as the median position and redshift of the member galaxies.
Note that some extremely large structures are not marked, because their median redshifts are out of range of the narrow redshift slice in Fig. \ref{newfig3}.

The blue cluster (See Section \ref{results}) actually has been identified by another cluster identification algorithm in a previous study \citep[SDSS-C4 3028;][]{Miller2005}. 
However the physical properties of this cluster, including the blue fraction, have not been focused on so far.
\\

\subsection{Physical properties of galaxy clusters}
\label{physical_property}
Once we identified the clusters and member galaxies, we measured the physical properties of the clusters such as the halo mass, velocity dispersion, virial radius, and crossing time. 
Following the literature \citep{Crook2007}, the halo mass of each cluster was calculated as
\begin{equation}
\label{eq5}
M_{\rm halo}=\frac{3\pi}{2}\frac{\sigma_{P}^{2}R_{p}}{G},
\end{equation}  
where G is the gravitational constant, $\sigma_{P}$ is the projected velocity dispersion 
\begin{equation}
\label{eq6}
\sigma_{P}^{2}=\frac{\Sigma_{i}(V_{i}-V_{c})^{2}}{N_{\rm member}-1},
\end{equation}
and $R_{P}$ is the projected virial radius,
\begin{equation}
\label{eq7}
R_{P}=\frac{N_{\rm member}(N_{\rm member}-1)}{\Sigma_{i>j}R_{ij}^{-1}}.
\end{equation}
$V_{c}$ is the recessional velocity of the cluster that was calculated by the median of recessional velocities of member galaxies, $V_{i}$.
$R_{ij}$ is the projected distance between $i$th and $j$th member galaxies.
The number of member galaxies is denoted by $N_{\rm member}$.

The cluster crossing time \citep{Ai2018}, $t_{c}$, is calculated as
\begin{equation}
\label{eq8}
t_{c}=\frac{1.51^{1/2}R_{\rm rms}}{3^{1/2}\sigma_{P}},
\end{equation}
where $R_{\rm rms}$ is the rms projected cluster radius \citep{Berlind2006} given by,
\begin{equation}
\label{eq9}
R_{\rm rms}=\sqrt{\frac{1}{N_{\rm member}}\Sigma_{i} r_{i}^{2}}.
\end{equation}
$r_{i}$ is the projected distance of $i$th member from the centre of the cluster.
The centre was derived from the median position of the member galaxies.\\

\subsection{Dressler-Shectman test}
We applied the Dressler-Shectman (DS) test \citep{Dressler1988,Ai2018} to the blue cluster (see Section \ref{results}) to check if the cluster has sub-structure.
Clearly relaxed systems will have less sub-structures.
If the halo is a relaxed system without substructure, the $\Delta$/N$_{\rm member}$ value is expected to be smaller than 1.0.
For the $i$th member galaxy, 10 nearest neighbours are used to derive a mean velocity ($\nu_{l,i}$) and velocity dispersion ($\sigma_{l,i}$).
We calculated $\delta_{\rm DS,i}$ for each member galaxy defined as 
\begin{equation}
\label{eq10}
\delta_{\rm DS,i}^{2}=\frac{11}{\sigma_{\rm cluster}^{2}}[(\nu_{l,i}-\nu_{\rm cluster})^{2}+(\sigma_{l,i}-\sigma_{\rm cluster})^{2}],
\end{equation}
where $\sigma_{\rm cluster}$ is the cluster velocity dispersion and $\nu_{\rm cluster}$ is the cluster mean velocity.
Based on Eq.\ref{eq10} we computed $\Delta$/N$_{\rm member}$, where $\Delta = \Sigma_{i}\delta_{\rm DS,i}$.
The $\Delta$/N$_{\rm member}$ value of the blue cluster is 0.89 which suggests a virialised system.\\

\subsection{Definition of two galaxy populations}
\begin{figure}
	\includegraphics[width=\columnwidth]{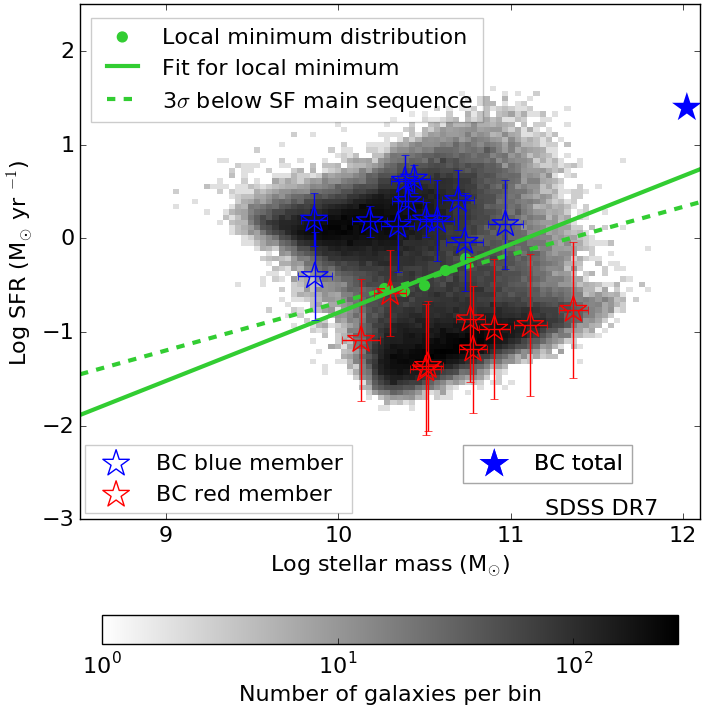}
    \caption{
    The boundary between blue star-forming and red quiescent galaxies in the SDSS DR7.
    The grey scale shows the number density of galaxies on this plot.
    The local minimum densities of five stellar-mass bins are marked by the green dots.
    The stellar mass bins are chosen to have a clear local minimum in density.
    The green solid line is the boundary adopted in this paper, which is defined by a linear fit to the local minima.
    The green dashed line is 3$\sigma$ below the blue main sequence 
    determined by a linear fitting of the local maxima of the SFR distributions in each stellar-mass bin.
    }
    \label{newfig4}
\end{figure}

We used the SFR and stellar mass of galaxies to divide them into two different populations. 
The \lq\lq main sequence\rq\rq\ of star-forming galaxies has been well explored so far in the stellar mass-SFR plane.
The blue star-forming galaxies on the main sequence can be easily distinguished from the red quiescent galaxies.
In Fig. \ref{newfig4}, we confirmed the clear bimodality of the galaxy distribution in our sample. 
To define the boundary of the bimodality, we divided our sample into five subsamples with different stellar-mass bins from $\log M_{*} = $10.2 to 10.8, so that the local minimum of the SFR distribution can be clearly determined.
In each mass bin, the local minimum of the $\log$SFR distribution is estimated and shown by green dots in Fig. \ref{newfig4}.
The boundary line is determined by a linear fitting of the local minima of the SFR distributions in each bin.
Here, we define galaxies above/below the boundary as blue star-forming/red quiescent galaxies, respectively.

In Fig. \ref{newfig4}, member galaxies of the blue cluster are also over-plotted (blue/red open stars) together with the total stellar mass and SFR of the blue cluster (filled blue star)\\


Here we briefly mention a blue fraction we derive for Abell 1367 in our analysis, because Abell 1367 is a well-known local cluster that shows a high fraction of spiral galaxies.
The redshift of Abell 1367 is 0.0213 \citep{Butcher1984}.
This is almost at the edge of our redshift selection criterion, i.e., $0.02 < z < 0.082$.
Therefore, about half of its member galaxies could be missed from our analysis. 
In fact, among sub-structures of Abell 1367, the only North West dense structure \citep{Cortese2004} was identified by our analysis. 
We determine a blue fraction of 0.15$\pm$0.06 for the North West dense structure of Abell 1367.
\citet{Butcher1984} estimated a spiral fraction of 0.19$\pm$0.03 for Abell 1367 as a whole. 
They used a morphological definition for spiral galaxies and an absolute magnitude limit of M$_{V}$ = -20 mag.
\citet{Cortese2004} investigated number fractions of emission-line galaxies in different sub-structures of Abell 1367 based on a flux-limited sample. 
They divided the sample into emission-line and non emission-line galaxies, and estimated the fraction of emission-line galaxies at the North West dense structure to be 0.31. 

\subsection{Semi-analytic Models and Simulation}
\label{SAM}
To compare with the SDSS galaxies, we used two semi-analytic models of galaxy formation and evolution: the GALACTICUS \citep{Benson2012} galaxy catalogue constructed in the framework of the MULTIDARK \citep{Knebe2018} cosmological simulation of dark matter halos and the semi-analytic galaxy catalogue \citep{Makiya2016} of the $\nu^{2}GC$ simulation \citep{Ishiyama2015}.
The volumes of these semi-analytic models (1 $h^{-1}$ Gpc$^{3}$ for GALACTICUS and 560 $h^{-1}$ Mpc$^{3}$ for the $\nu^{2}$GC-M galaxy catalogue) are large enough to minimise statistical uncertainty when the physical properties of the galaxy clusters are compared with the SDSS clusters.

\begin{figure*}
	\includegraphics[width=2.25in]{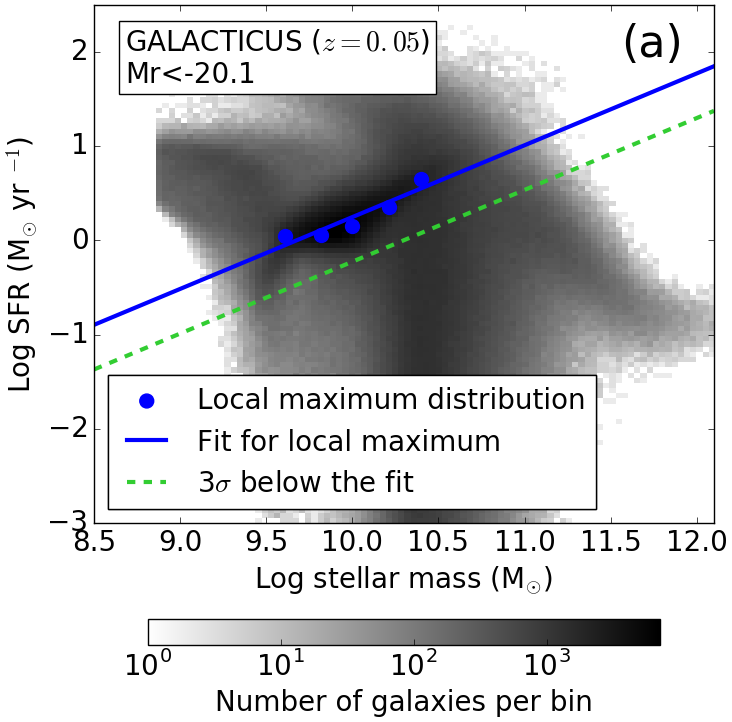}
	\includegraphics[width=2.25in]{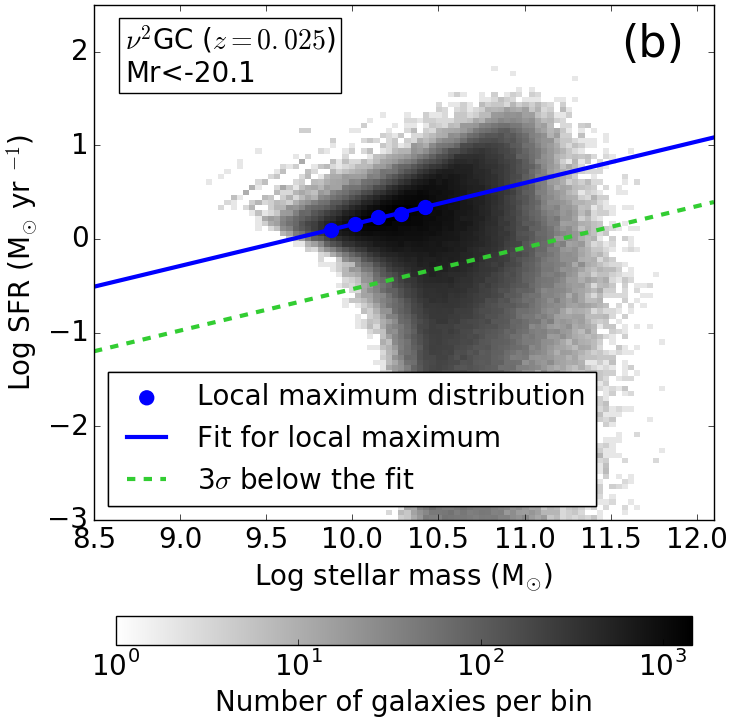}
	\includegraphics[width=2.25in]{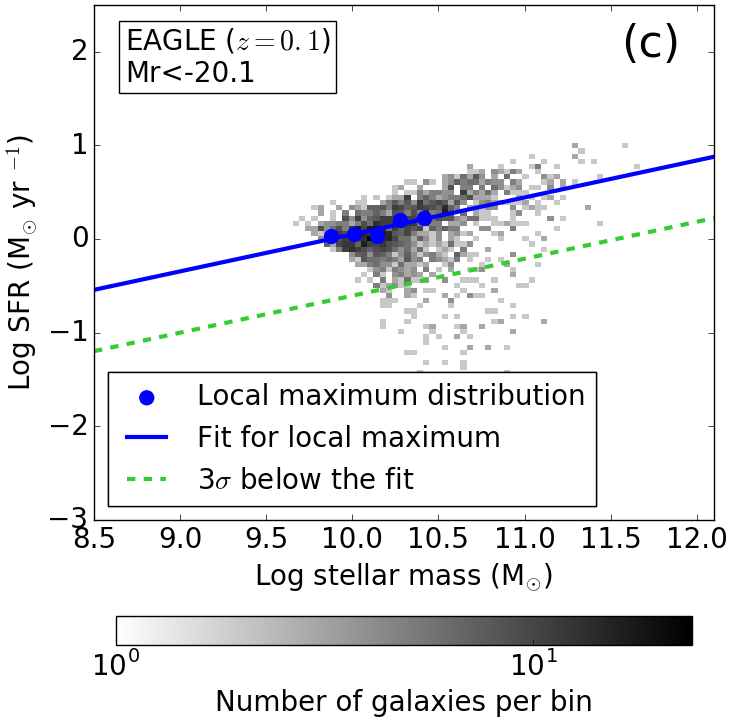}
    \caption{
    The boundary between blue star-forming and red quiescent galaxies in the semi-analytic models and the EAGLE simulation.
    The magnitude-matched samples are shown (see Section \ref{SAM} for details).
    The grey scale shows the number density of galaxies on this plot.
    The local maximum densities of five stellar-mass bins are marked by the blue dots.
    The stellar mass bins are chosen to have a clear local maximum density.
    The blue solid line is the blue main sequence defined by a linear fit to the local maxima.
    The green dashed line is the boundary adopted in this paper, which is defined as 3$\sigma$ below the blue main sequence (blue solid line).
    }
    \label{newfig5}
\end{figure*}

In the MULTIDARK simulation, three different semi-analytic models are provided, namely, SAG, SAGE, and GALACTICUS. 
Each model matches some of the observed empirical relations but not others.
SAG's strength is in providing reasonable gas fractions and metallicity relations. 
SAGE can better reproduce the stellar mass function of galaxies and stellar-to-halo mass relations. 
GALACTICUS better matches 
the SFR function and its evolution. 
Here, our major concern is the number fraction of blue star-forming galaxies in the cluster environment.
It is likely more directly linked to the star-formation rate function of galaxies (better in GALACTICUS), rather than the other advantages in SAG and SAGE.
Actually galaxies in GALACTICUS show the colour bimodality correctly \citep{Knebe2018}. 
Therefore, we selected the GALACTICUS model in the MULTIDARK simulation at the snapshot of $z=0.05$. 
Although GALACTICUS shows colour bimodality, it does not perfectly match with that of observed galaxies and is not clear enough to define a local minimum density in the stellar mass-SFR plane (Fig. \ref{newfig5}a). 

In the $\nu^{2}$GC-M simulation, the mass resolution of dark matter halo, $2.2\times 10^{8}$ M$_{\odot}$, is much better than that of GALACTICUS (1.5$\times 10^{9}$ M$_{\odot}$), while the volume is still large enough for our purpose.
The semi-analytic model of $\nu^{2}$GC includes fundamental physics of galaxy formation and evolution such as gas cooling, photoionization heating due to an UV background, star formation/feedback, growth of super massive black hole, and AGN feedback.
We used $\nu^{2}$GC galaxies at the snapshot of $z=0.025$. 
Although $\nu^{2}$GC successfully reproduces the observed slope of the main sequence and distribution of star-forming galaxies in the stellar mass-SFR plane \citep{Makiya2016}, $\nu^{2}$GC does not reproduce a bimodal galaxy distribution (Fig. \ref{newfig5}b).
The method we use to define the boundary of two populations in SDSS is therefore not applicable to $\nu^{2}$GC.

In addition to the two semi-analytic models, we compared with the galaxies from the EAGLE project \citep{Schaye2015,Crain2015}, which is a suite of hydrodynamical simulations that follow the formation of galaxies and supermassive black holes in cosmologically representative volumes of a standard $\Lambda$ cold dark matter universe. 
We used the fiducial model of RefL0100N1504, which is in broad agreement with the observed colour distribution of the local galaxies at $z=0.1$ \citep{Trayford2015}, although the volume is relatively small compared with that of the semi-analytic models we compare with.
The galaxies in the EAGLE simulation show the bimodality in the stellar mass-SFR plane.
However, 
many red quiescent galaxies have zero SFR and are therefore not shown in Fig. \ref{newfig5}c.

In all of the models, the local minimum density cannot be defined clearly.
Therefore, we applied a similar, but different method to the GALACTICUS, $\nu^{2}$GC, and EAGLE to determine the boundary between blue star-forming and red quiescent galaxies. 
First, we traced the peaks of SFR distributions in subsamples corresponding to different stellar mass bins (blue dots in Fig. \ref{newfig5}).
The stellar mass bins are selected so that the peaks of SFR distributions can be clearly determined.
These peaks are fitted by a linear function to define a main-sequence line.
Secondary, we examined how widely star-forming galaxies distribute around the main-sequence line.
The standard deviation ($\sigma$) of data offset from the main-sequence line is calculated based on only the galaxies above the main-sequence line to avoid the contamination from red quiescent galaxies to the data dispersion.
Here, we fit the offset distribution with a Gaussian function to derive $\sigma$.
Finally, we defined the boundary line (green dashed line in Fig. \ref{newfig5}) as the 3$\sigma$ offset from the main-sequence line with the same slope. 
We checked this method by using the SDSS sample in Fig. \ref{newfig4}. 
The boundary defined to be 3$\sigma$ away from the main-sequence line (green dashed line in Fig. \ref{newfig4}) is almost identical to that defined by the local minima of SFR distributions (green solid line in Fig. \ref{newfig4}).

In all of the models, the SFR, stellar mass, position (X,Y,Z), peculiar velocity ($V_{X}$,$V_{Y}$,$V_{Z}$), and $r$-band absolute magnitude of galaxy (M$_{r}$) are provided. 
To perform exactly the same analysis as that we have done for SDSS galaxies, we converted the physical distance 
along the X axis to redshift assuming the {\it Planck15} \citep{Planck2015} cosmological parameters.
Here, X=0 corresponds to redshifts of 0.05, 0.025, and 0.1 for GALACTICUS, $\nu^{2}$GC, and EAGLE, respectively.
The peculiar velocity along the line of sight ($V_{X}$) was also included in the redshift calculation.
The volume-limited samples from the models were constructed by selecting galaxies brighter than M$_{r}^{*}$+1.5 = -20.1 mag in the SDSS $r$-band.
This is the same as our SDSS selection, i.e., a magnitude-matched samples.
Based on the SFR, stellar mass, projected coordinate (Y,Z), and redshift of the volume-limited samples, we simulated \lq \lq observations\rq \rq\ by applying the same analysis, cluster identification algorithm, and Eqs. \ref{eq5}-\ref{eq9}. 

In addition to the magnitude-matched samples mentioned above, we constructed density-matched samples from GALACTICUS, $\nu^{2}$GC, and EAGLE.
We used a M$_{r}^{*}$+1.15 mag limit for GALACTICUS, M$_{r}^{*}$+1.2 for $\nu^{2}$GC, and M$_{r}^{*}$+1.4 for EAGLE, which results in samples with mean density equivalent to that of the SDSS DR7 sample with a M$_{r}^{*}$+1.5 mag limit.
Because we adopt an identical linking length of 0.75 Mpc, the density-matched samples ensure equal numbers of member galaxies for halos of the same mass in the different models and the SDSS sample.
For the density-matched samples, we performed all the analyses described in Section \ref{analysis} in the same manner as the magnitude-matched samples.
\\

\subsection{Accuracy of halo mass estimate}
Since the semi-analytic models provide the true host halo mass for individual galaxies, the accuracy of our halo mass estimate can be examined by comparing with the true halo mass ($M_{\rm halo, true}$) and mass derived from our cluster algorithm and Eq. \ref{eq5} ($M_{\rm halo,derived}$).
Individual member galaxies identified by our algorithm have $M_{\rm halo, true}$ that are assigned in the models \citep{Benson2012,Makiya2016}.
Our algorithm sometimes links together galaxies belonging to multiple independent halos into a single cluster; in such cases we adopt the maximum halo mass among the constituent galaxies as $M_{\rm halo, true}$.

GALACTICUS and $\nu^{2}$GC use different definitions of $M_{\rm halo, true}$.
While GALACTICUS adopts $M_{\rm 200c}$, i.e., halo mass within a radius where the averaged cluster mass density is 200 times the critical density of the Universe, $\nu^{2}$GC is based on tje total mass of dark matter particles identified by a friends-of-friends method, $M_{\rm FoF}$. 
$M_{\rm FoF}$ could be systematically larger than $M_{\rm 200c}$ \citep[e.g.,][]{More2011}.
$M_{\rm 200c}$ should be close to the virial masses calculated by Eq. \ref{eq5}. 
Therefore we converted $M_{\rm FoF}$ of $\nu^{2}$GC to $M_{\rm 200c}$ by applying the best-fit function between $M_{\rm FoF}$ and $M_{\rm 200c}$ to obtain $M_{\rm halo, true}$ (see APPENDIX A for details).

\begin{figure}
	\includegraphics[width=\columnwidth]{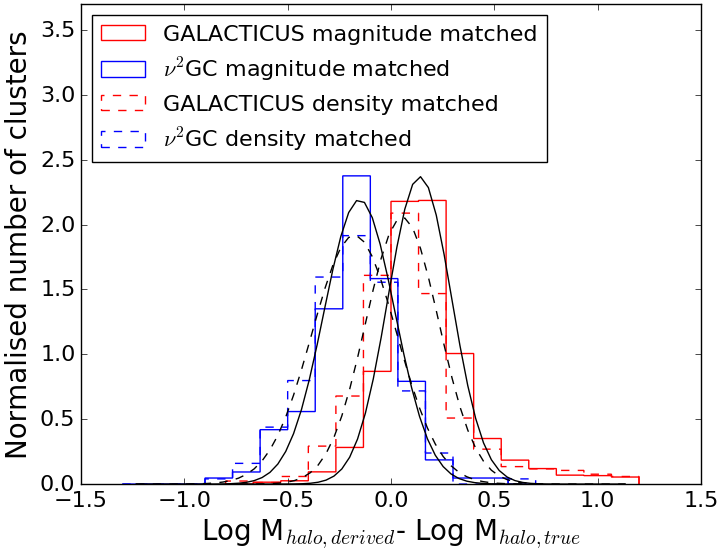}
    \caption{
    Uncertainty of halo mass estimate.
    The normalised histograms of the differences between the true halo mass and derived halo mass through Eq. \ref{eq5} for GALACTICUS (red) and $\nu^{2}$GC (blue) model clusters.
    Different line styles correspond to magnitude-matched and density-matched samples (see Section \ref{SAM} for details).}
    The best-fit Gaussian functions are shown with black solid/dashed lines.
    \label{newfig6}
\end{figure}

The difference between $M_{\rm halo,derived}$ and $M_{\rm halo,true}$ is shown in Fig. \ref{newfig6}.
Two histograms correspond to GALACTICUS and $\nu^{2}$GC, which are consistent within 1 $\sigma$.
We found that $\log$($M_{\rm halo, derived}$) is overestimated by 0.14 dex with $\sigma=0.15$ in GALACTICUS and is underestimated by $-0.15$ dex with $\sigma=0.17$ in $\nu^{2}$GC.
Taking these systematics together we estimated the uncertainty of halo mass to be $\log$($M_{\rm halo, derived}$)$^{+0.29}_{-0.32}$ which corresponds to $2.0^{+1.9}_{-1.0} \times 10^{14}$ M$_{\odot}$ for the blue cluster.
In Fig. \ref{newfig6}, we confirmed that the difference between magnitude-matched and density-matched samples is negligible in terms of halo mass estimate.

\section{Results}
\label{results}
\begin{figure*}
	\includegraphics[width=7.0in]{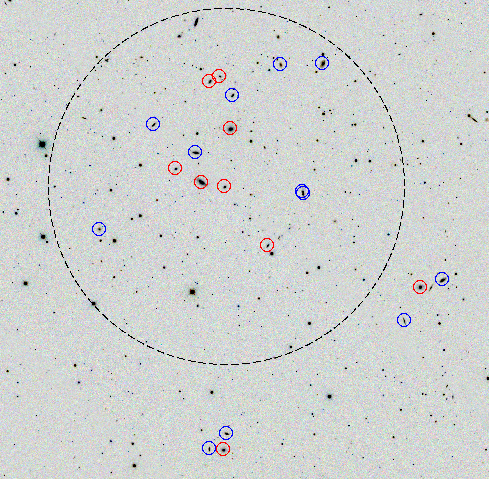}
    \caption{
    The blue cluster at ${\bf z=0.061}$ found in the SDSS DR7.
    The member galaxies classified as the blue star forming are marked by blue circle, while the red circle for the red quiescent galaxies.
    The estimated virial radius of the blue cluster, 0.64 Mpc (8.7 arcmin on sky), is shown with a black dashed circle.
    The centre of the black dashed circle is determined by the median J2000.0 coordinates of the member galaxies, i.e., (RA, Dec) = (09:07:38.83, 52:04:04.09).
    The field of view is 1.76 $\times$ 1.76 Mpc.
    }
    \label{newfig7}
\end{figure*}

\begin{figure*}
	\includegraphics[width=7.0in]{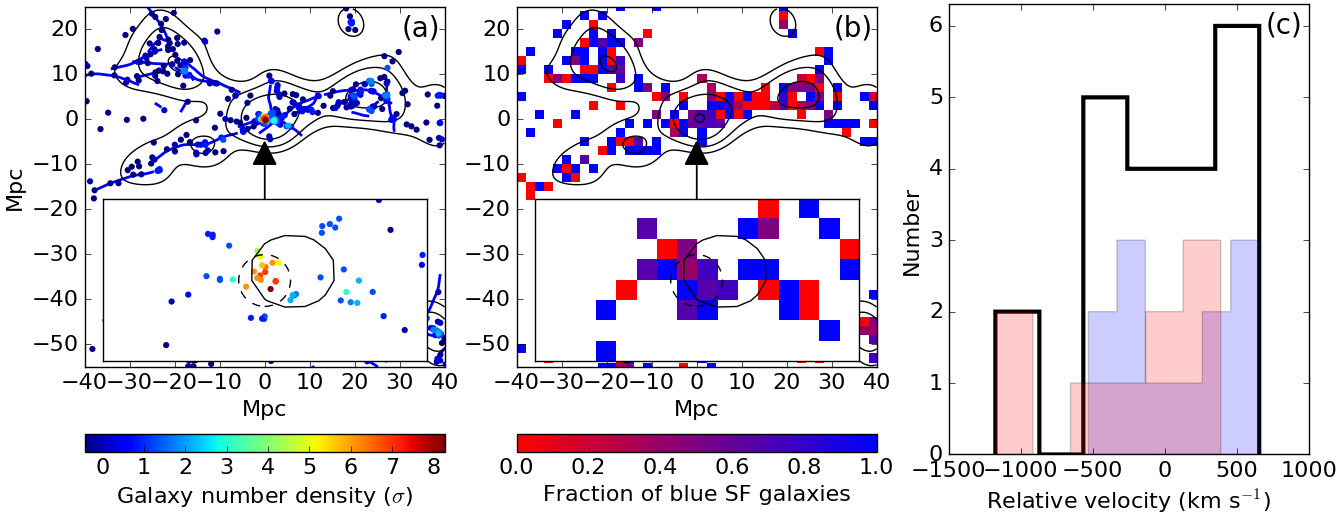}
    \caption{
    The environment of the blue cluster.
    (left) The projected number density of galaxies around the blue cluster. 
    Individual galaxies within $\pm$ 1000 km s$^{-1}$ around the redshift of the blue cluster are shown by dots.
    The density is calculated within the linking length of 0.75 Mpc radius and $\pm$ 1000 km s$^{-1}$ around individual galaxies in unit of $\sigma$, where $\sigma$ is the standard deviation of the projected density measured from the median density.
    Contours correspond to median, 1.0$\sigma$, 3.0$\sigma$, and 8.0$\sigma$ densities approximated by the Gaussian kernel density estimation.
    Blue solid lines are projected filamentary structures identified by \citet{Tempel2014} within $\pm$ 1000 km s$^{-1}$ from the redshift of the blue cluster.
    The inner box is 8.0$\times$4.0 Mpc region around the blue cluster.
    The virial radius of the blue cluster, 0.64 Mpc, is indicated by a black dashed circle, while the 8.0$\sigma$ density contour is displayed by a solid contour.
    The blue and red colours correspond to the median and $\sim$8.0$\sigma$ densities, respectively.
    (middle) Same as the left panel except for the fraction of the blue star-forming galaxies, $f_{b}$, calculated in each 2.0 $\times$ 2.0 Mpc$^{2}$ bins.
    The inner box is binned by 0.5 $\times$ 0.5 Mpc$^{2}$.
    The blue and red colours correspond to $f_{b}$=1.0 and 0.0, respectively.
    (right) Projected relative velocity distribution of the member galaxies in the blue cluster.
    Black, blue, and red histograms are velocity distributions of total member, blue star-forming, and red quiescent galaxies, respectively.
    }
    \label{newfig8}
\end{figure*}

In this work, we discovered a galaxy cluster which show an unusually large faction of blue star-forming galaxies at redshift $z$=0.061 composed of 21 member galaxies (Fig. \ref{newfig7} for image and Fig. \ref{newfig8}a for galaxy distribution). 
Physical properties, optical images, and spectra of 21 member galaxies are listed in appendix~B.
The median equatorial coordinates (J2000.0) of the member galaxies are (RA, Dec) = (09:07:38.83, +52:04:04.09).
The blue galaxy fraction of this cluster is 0.57$\pm$0.06 (Fig. \ref{newfig8}b), which is much higher than that of typical dense galaxy clusters with similar mass in the local Universe \citep{Butcher1984}.
The virial mass is estimated to be 2.0$^{+1.9}_{-1.0} \times 10^{14}$ M$_{\odot}$ based on the projected velocity dispersion of the member galaxies (Fig. \ref{newfig8}c), $\sigma_{P}$ = 510.0 km s$^{-1}$, and the projected virial radius, $R_{P}=0.64$ Mpc. 
The Dressler-Shectman (DS) test \citep{Dressler1988,Ai2018} implies that the cluster is already dynamically virialized ($\Delta$/N$_{\rm member}<$1.0, see Section \ref{analysis} for details).
The estimated halo mass is consistent with that from the empirical relationship between the velocity dispersion and halo mass calibrated by a weak lensing mass \citep{Hamana2009} (2.5$\times$10$^{14}$ M$_{\odot}$).
The galaxy density in the cluster is $\sim$ 8 $\sigma$ higher than the average density of galaixes in the field (Fig. \ref{newfig8}a).

Galaxy clusters including the blue cluster were identified in the SDSS DR7 \citep{Abazajian2009} as follows (see Section \ref{analysis} for details).
First, we selected galaxies in the dense environments, i.e., $\delta \geqq 8.0 \sigma$, where $\delta$ is the projected number density of galaxies within the linking length of 0.75 Mpc and $\pm$ 1000 km s$^{-1}$, and $\sigma$ is standard deviation of $\delta$.
Second, for each selected galaxy, member galaxies were identified based on the friends-of-friends method with a linking length of 0.75 Mpc and relative velocity within $\pm$ 1000 km s$^{-1}$.

\begin{figure*}
	\includegraphics[width=2.25in]{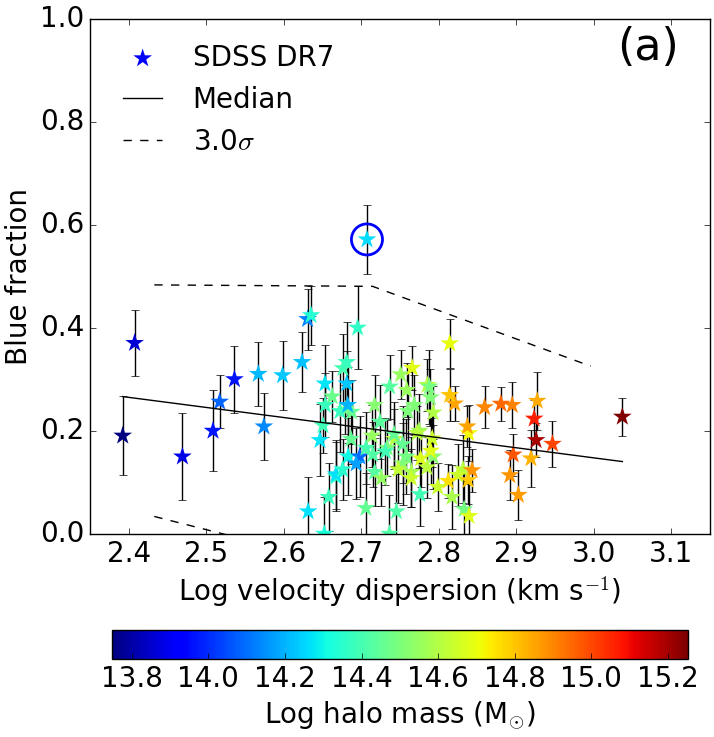}
	\includegraphics[width=2.25in]{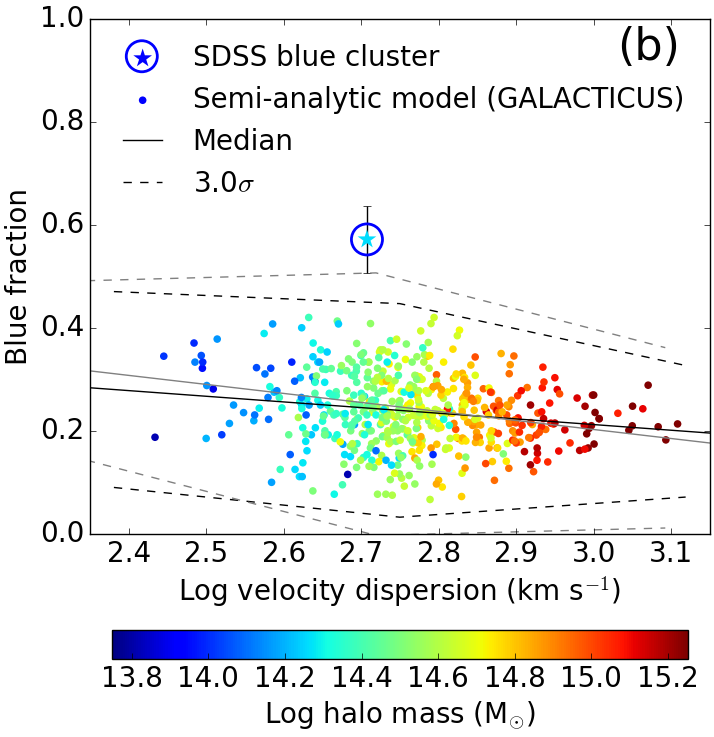}
	\includegraphics[width=2.25in]{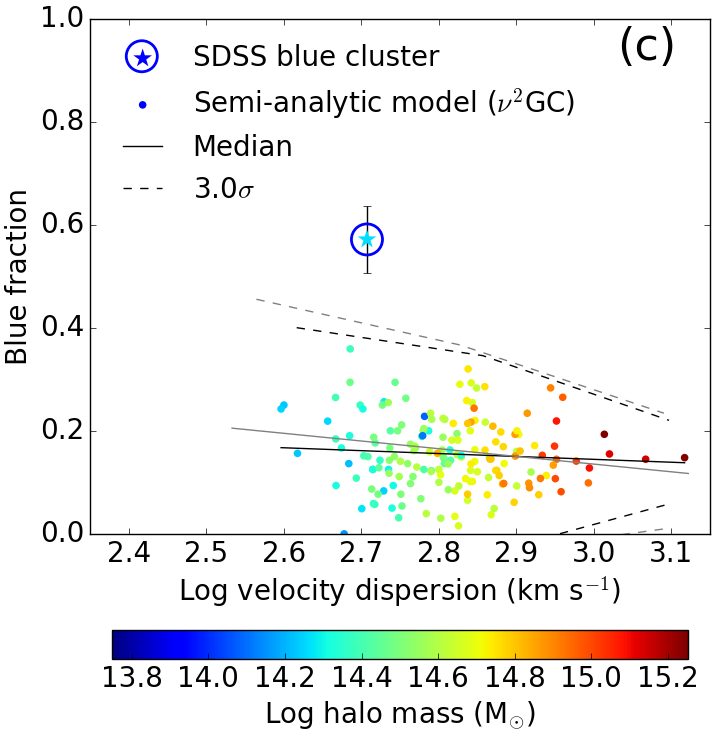}
    \caption{
    The fraction of the blue star-forming galaxies as a function of the cluster velocity dispersion.
    (left) The galaxy clusters identified in our analysis of the SDSS DR7 data.
    The errors of individual clusters are calculated by the Monte Carlo simulation on the galaxy main sequence plane, i.e., SFR as a function of the stellar mass, by using observational uncertainties of the SFR and stellar mass. 
    The blue cluster is demonstrated by the encircled star with the error bar.
    (middle) The same as the left panel but for galaxies calculated in the semi-analytic model of GALACTICUS at the snapshot of $z=0.05$.
    The magnitude-matched sample is shown.
    (right) The same as the left panel but for galaxies calculated in the semi-analytic model of $\nu^{2}$GC at the snapshot of $z=0.025$.
    The magnitude-matched sample is shown.
    In all panels, the black solid and dashed lines indicate the median lines of the data distributions and 3$\sigma$ deviations from them, respectively. 
    Grey solid and dashed lines in the middle and right panels are the medians and 3$\sigma$ deviations calculated from density-matched sample (see Section \ref{SAM} for details).
    The colours correspond to the halo mass estimated from the velocity dispersion and virial radius.
    All clusters in both of the SDSS and galaxy models are identified in the dense environment with $\delta \geq 8 \sigma$, where $\delta$ is the projected number density of galaxies within $\pm$ 1000 km s$^{-1}$.
    The error bars are calculated by Monte Carlo simulations including observational uncertainties of SFR and stellar mass.
    }
    \label{newfig9}
\end{figure*}

We confirmed the blue fraction decreases with increasing cluster velocity dispersion (Fig. \ref{newfig9}a), as expected. 
However, in Fig. \ref{newfig9}a, there is one cluster whose blue fraction is 4.0 $\sigma$ higher than the median for the SDSS clusters. 
The corresponding probability to find such a high blue fraction in an individual cluster is only 0.003\% assuming a gaussian distribution around the median. 
We also checked the likelihood of finding the blue cluster by performing a simple simulation as follows. 
First, we took all the member galaxies in the clusters identified by our analysis together. 
Secondary we shuffled the member galaxies and re-constructed clusters keeping the numbers of member galaxies in each cluster fixed. 
Based on the shuffled clusters, blue fractions were calculated. 
This process was iterated by 100,000 times. 
A number fraction of shuffled clusters with blue fractions of $f_{b} \geqq 0.57$ is 0.0032\%. 
This value is consistent with 0.003\% derived from 4 sigma deviation of the blue cluster. 
Since a total of 100 SDSS clusters are identified in our analysis, the probability of finding one or more blue clusters with $f_{b} \geqq 0.57$ among the full sample of 100 clusters is approximately $100 \times 0.0032 = 0.32$\%.

To see how unusual this is in the cold dark matter model with a cosmological constant ($\Lambda$CDM), the blue fraction was compared with the state-of-the-art semi-analytic models of the galaxy formation and evolution (Fig. \ref{newfig9}b and c). 
We selected GALACTICUS \citep{Benson2012} and $\nu^{2}$GC models \citep{Ishiyama2015,Makiya2016}, since these semi-analytic models have large volumes which provide sufficient number of clusters for statistical comparisons.
For a fair comparison, we simulated \lq \lq observations\rq \rq\ in the galaxy catalogues constructed by GALACTICUS ($z=0.05$) and $\nu^{2}$GC ($z=0.025$) projects.
The same selection criteria as in the SDSS DR7 is applied to these two catalogues. 
For model galaxies in environments with an overdensity of 8 $\sigma$ or higher, we used the friends-of-friends method to identify galaxy clusters with the same linking length and velocity criteria.
We note that the model calculations could have a difficulty in reproducing the physical properties of the small or faint galaxies in general.
The colour distribution or SFR of faint galaxies can deviate from the observed distributions. 
To avoid these problems, we adopted a bright magnitude cut, M$_{r}^*$+1.5=$-20.1$ mag, i.e., magnitude-matched samples (see Section \ref{SAM} for details).
At this magnitude cut, the distribution of the blue fractions in the two semi-analytic models are broadly consistent with the SDSS DR7 data. 
The semi-analytic models also show a negative correlation between the blue fraction and velocity dispersion, though the trend is weaker in the $\nu^{2}$GC model (Fig. \ref{newfig9}b and c).
The blue fraction of the SDSS blue cluster is still much higher than those of the semi-analytic models, i.e., 4.7$\sigma$ and 5.7$\sigma$ away from the median lines in GALACTICUS and $\nu^{2}$GC, respectively.
In Fig. \ref{newfig9}b and c, we confirmed that the difference between magnitude-matched and density-matched samples have no significant impact on the blue fractions.

Hydrodynamical simulations of galaxy formation can be more reliable in terms of the SFR and colour of faint galaxies thanks to their more sophisticated assumptions or approximations of the underlying physics.
Therefore, we also compared the blue cluster in the SDSS with the EAGLE \citep{Schaye2015,Crain2015} simulation of the galaxy formation.
The fiducial model of RefL0100N1504 selected here has the largest volume in the EAGLE simulations. However, this is still only $\sim$2\% of the survey volume of the SDSS DR7. 
The expected number of clusters with our method is only a few.
Therefore a statistical approach is not feasible due to such a small sample.
Since the redshift snapshots of the EAGLE simulations do not overlap with the redshift range of our SDSS sample, we used a snapshot at a higher redshift, $z=0.1$, for a conservative estimate of blue fractions.
Among four clusters we identified in the EAGLE simulation, the highest blue fraction is 0.48 for both cases of magnitude-matched and density-matched samples, which is still lower than that of the blue cluster in the SDSS.\\

\section{Discussion}
\label{discussion}
Neither the semi-analytic models nor the EAGLE simulation predict a blue fraction as high as that of our blue cluster.
The deviation of 4.0 $\sigma$ from the median blue fraction of the SDSS clusters is statistically significant. The significance is even greater when the blue cluster is compared with theoretical models ($4.7\sigma$ for GALACTICUS and 5.7$\sigma$ for $\nu^{2}$GC). The blue cluster is unlikely to be due to the stochastic variation of clusters.
The blue cluster could hint at physics that is not included in the models and simulations.\\

\subsection{Dynamical age of the cluster}
The blue cluster could be undergoing a very early stage of cluster evolution, even though it is in the local Universe. 
The DS test implies it is a relaxed system, i.e., $\Delta/{\rm N_{\rm member}} < 1.0$. 
A virialised cluster galaxy population, in which the orbits are random and isotropic, should show a Gaussian velocity distribution \citep[e.g.,][]{Merritt1987,Colless1996,Mamon2019}.
A simple test for Gaussianity is to measure the kurtosis of the velocity distribution \citep[e.g.,][]{Haines2015}, where the kurtosis of Gaussian distribution corresponds to 0.0. 
\citet{Haines2015} reported that the velocity distributions of star-forming galaxies in nearby clusters typically have kurtosis $\lesssim$ -0.6, closer to a top-hat profile than a Gaussian.
This suggests an infalling motion of star-forming galaxies in the clusters. 

The derived kurtosis for the blue cluster is -0.19, implying no clear evidence of non Gaussianity as a whole member galaxies. 
Figure 8 (c) shows the velocity distributions of blue star-forming and red quiescent galaxies in the blue cluster. 
Except for two red galaxies at $\sim$-1000 km s$^{-1}$, blue star-forming galaxies have relatively higher velocity offset from the velocity centre than red quiescent galaxies. 
This might suggest an infalling population of blue star-forming galaxies, although the number of sample is too small for a reliable conclusion.

\begin{figure*}
	\includegraphics[width=2.25in]{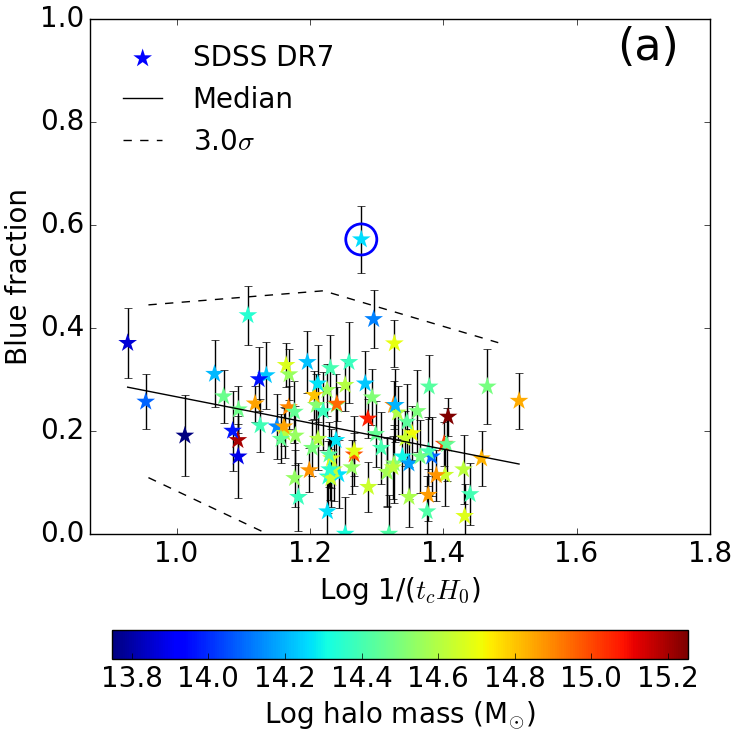}
	\includegraphics[width=2.25in]{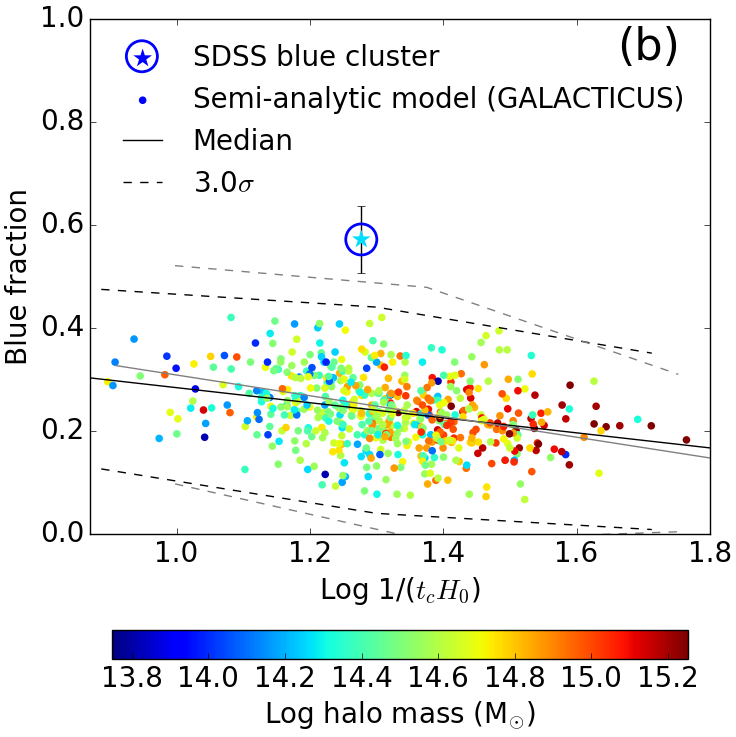}
	\includegraphics[width=2.25in]{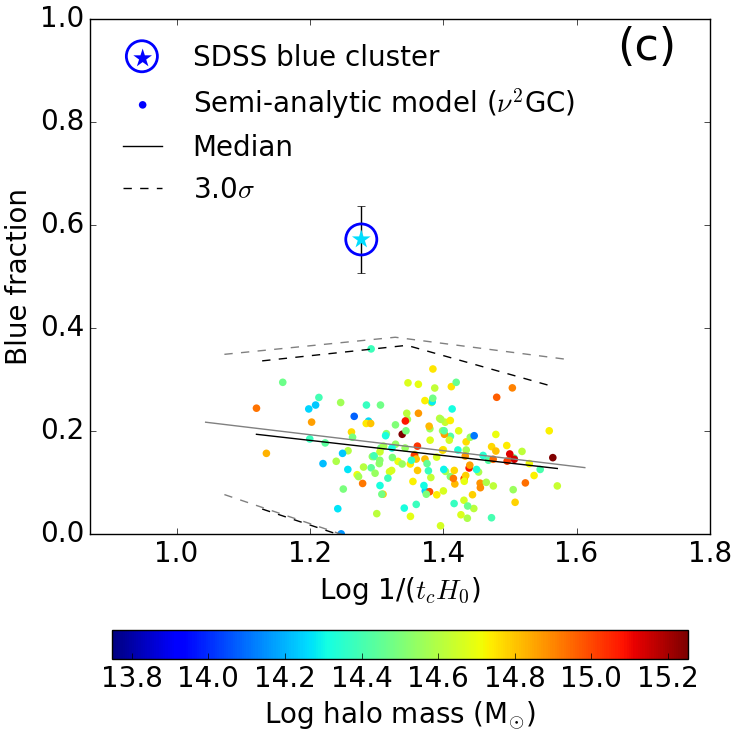}
    \caption{
    The fraction of the blue star-forming galaxies as a function of $1/(t_{c}H_{0})$.
    Same as Fig. \ref{newfig9} except for 1/($t_{c}H_{0}$) in the horizontal axis, where $1/(t_{c}H_{0})$ is the rough number of times a member galaxy could have traversed the cluster since its formation, i.e., an indicator of how much the system is dynamically relaxed.
    }
    \label{newfig10}
\end{figure*}

The cluster crossing time, $t_{c}$, is an age indicator for clusters \citep{Ai2018}.
The parameter $1/(t_{c}H_{0})$, where $H_{0}$ is the Hubble constant, tells us the extent to which the system is dynamically relaxed. 
This parameter is a rough estimate of how many times a galaxy could have traversed the cluster since its formation. 
A dynamically more relaxed system is expected to show both a higher value of $1/(t_{c}H_{0})$ and a lower blue fraction. 
We confirmed this in both the SDSS data and semi-analytic models (Fig. \ref{newfig10}), and there is also evidence in the literature \citep{Ai2018} that dynamically more relaxed clusters show a lower blue fraction. 
The blue cluster is located in the middle of the distribution of $1/(t_{c}H_{0})$ (Fig. \ref{newfig10}a), implying that it is not at a very early stage of dynamical evolution, despite its high blue fraction (Fig. \ref{newfig10}b, and c).
According to cosmological N-body simulations in $\Lambda$CDM, the typical formation epoch of a halo with 2.0$\times$10$^{14}$ M$_{\odot}$ is $z\sim$0.6, at which time the halo has accumulated half of its current mass \citep{Ishiyama2015}.
Under the assumption that the blue cluster is hosted by a typical halo of this mass, its formation epoch was $\sim$ 5.2 Gyr ago.
This time scale is comparable to or longer than the typical depletion timescale of molecular gas in star-forming galaxies \citep[e.g., $\sim$ 4 Gyr;][]{Peng2015}. 
In addition, there could be environmental effects that accelerate the transition from blue to red galaxy colours. 
Thus, the blue fraction is not expected to be as large as 0.57 in such a situation.
This argues against a young dynamical age as the explanation for the high blue fraction of the cluster.
Observational evidence suggest that the cluster is dynamically evolved and member galaxies have had a sufficient time to exhaust the gas, aided by environmental effects in the dense environment.
However, the galaxy population is curiously very blue.\\

\subsection{Stellar mass function}
\begin{figure*}
	\includegraphics[width=3.45in]{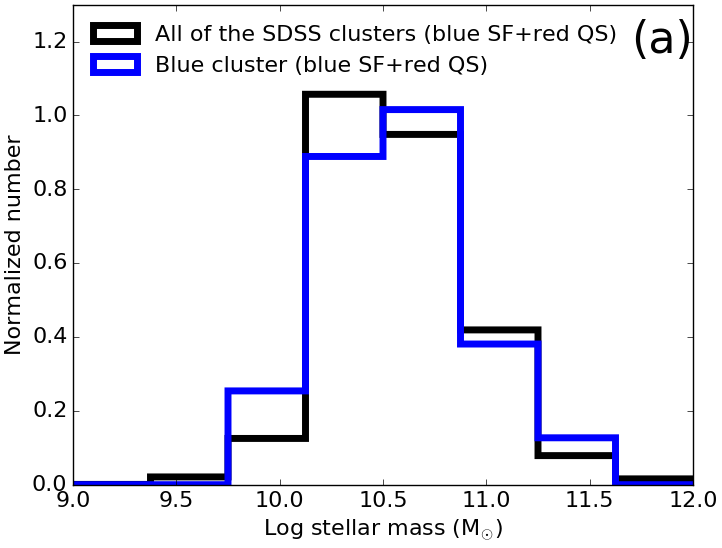}
	\includegraphics[width=3.45in]{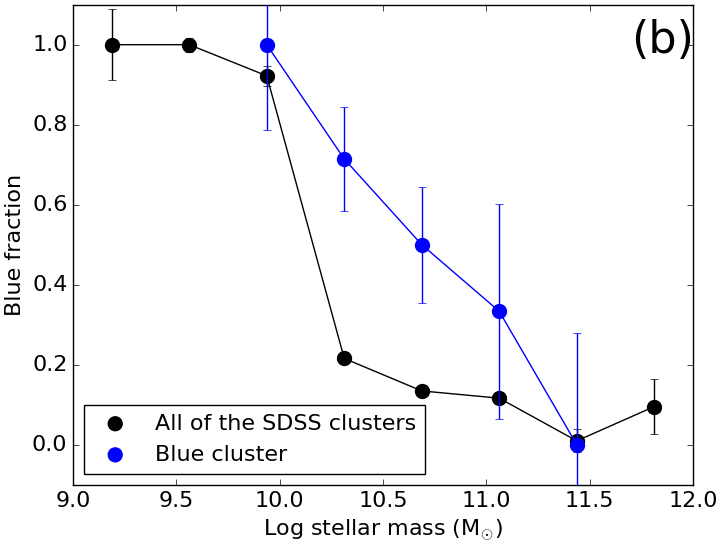}
    \caption{
    The galaxy stellar mass functions and blue fraction as a function of stellar mass.
    (left) The black open histogram is the normalised galaxy stellar mass function of all cluster member galaxies identified in the SDSS DR7. 
    The normalised galaxy stellar mass function of the blue cluster member galaxies is indicated by the blue open histogram. 
    (right) The fractions of blue star-forming galaxies of all the cluster member galaxies and the blue cluster are shown by black and blue lines, respectively.
    The error bars are calculated by Monte Carlo simulations including observational uncertainties in SFR and stellar mass.
    }
    \label{newfig11}
\end{figure*}

The galaxy stellar mass function might provide a hint to understand the high blue fraction. 
Since less massive galaxies show a higher blue fraction in general, it is possible that the blue cluster hosts an unusually large number of less massive galaxies that can contribute to its high blue fraction. 
The galaxy stellar mass function for all cluster member galaxies identified in the SDSS DR7 was compared with that of the blue cluster (Fig \ref{newfig11}a).
We found no clear difference between the two. 
The p-value of a Kolmogorov-Smirnov test is 0.94, indicating no significant difference between the two samples. 
Thus, it is unlikely that there is an excess of less massive galaxies in the blue cluster.
The blue fraction of the blue cluster at a fixed galaxy stellar mass is significantly higher than the average fraction between $10^{10}$ and $10^{11}$ M$_{\odot}$ (Fig. \ref{newfig11}b). 
This suggests that individual galaxies already have a relatively higher SFR at a fixed galaxy stellar mass, i.e., higher specific SFR (sSFR; defined as SFR/M$_{*}$, where M$_{*}$ is the galaxy stellar mass).
\\

\subsection{High sSFR and mergers}

We investigated the sSFR averaged across the individual clusters, calculated as SFR$_{\rm total}$/M$_{\rm *, total}$, where SFR$_{\rm total}$ and M$_{\rm *,total}$ are the total SFR and M$_{*}$ of the member galaxies in each cluster.
The sSFR of the blue cluster is the highest among our SDSS clusters and is consistent with the star-forming regime (Fig. \ref{newfig12}a and \ref{newfig4}). 
One possible explanation for the high sSFR is 
a higher frequency of major mergers than an average frequency in the cluster.
Major mergers may trigger intense star formation, hence high sSFR \citep{Sanders1988,Barnes1991}.
We checked the morphology of the member galaxies of the blue cluster (Fig. \ref{newfig7} and zoom-in images in Fig \ref{newfigB1}). 
Only one pair of the member galaxies is undergoing a major merger by the visual inspection.
While the implied merging fraction of $\sim$ 10\% is slightly higher than that of typical field \citep{Darg2010} and cluster environments \citep{Dokkum1999,McIntosh2008} in the local Universe, there is no clear evidence of an especially high merging fraction in the blue cluster.
Since other blue star-forming galaxies do not show any signs of merging, major mergers are unlikely to be the main driver of high sSFR and blue fraction in the blue cluster.\\

\subsection{High sSFR and gas mass fraction}

\begin{figure*}
	\includegraphics[width=2.25in]{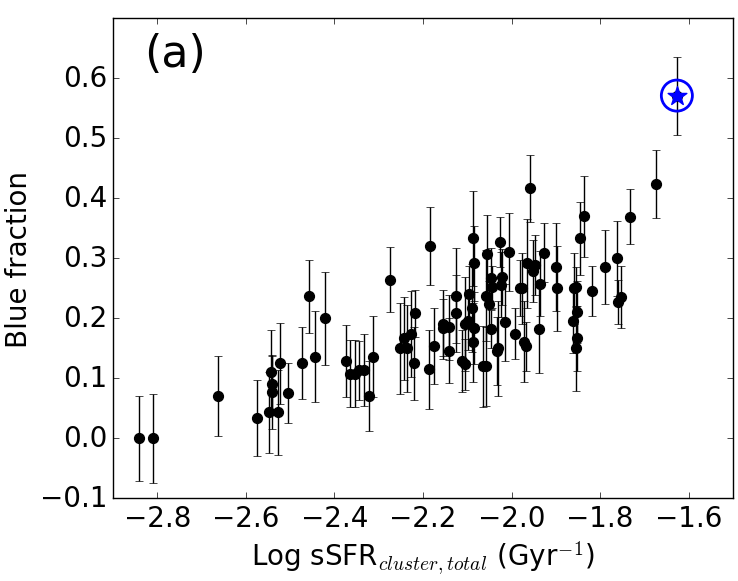}
	\includegraphics[width=2.25in]{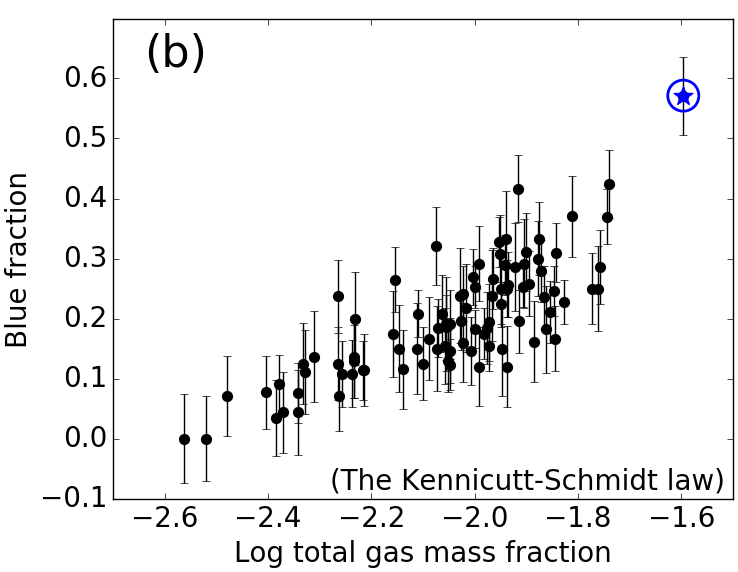}
	\includegraphics[width=2.25in]{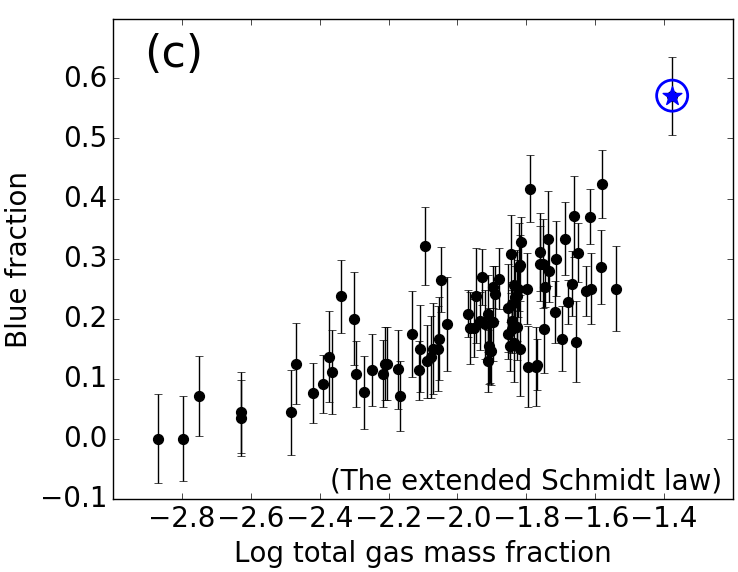}
    \caption{
    The fraction of blue star-forming galaxies as a function of sSFR and gas mass.
    (left) The blue fraction of the SDSS DR7 clusters as a function of the sSFR (=SFR/M$_{*}$).
    The blue cluster is indicated by the encircled blue star.
    (middle) Same as left except for the gas mass fraction (=M$_{gas}$/M$_{*}$) in the horizontal axis, which is estimated from the Kennicutt-Schmidt law \citep{Kennicutt1998}.
    (right) Same as middle except for the extended Schmidt law \citep{Shi2018}.
    The error bars are the same as Fig. \ref{newfig9}.
    }
    \label{newfig12}
\end{figure*}

The high sSFR in the blue cluster suggests the presence of a large amount of gas in its member galaxies, according to the Kennicutt-Schmidt law, which is the empirical relationship between SFR surface density, $\Sigma_{\rm SFR}$ (=SFR/$\pi r^{2}$), and gas mass surface density, $\Sigma_{M_{\rm gas}}$ (=$M_{\rm gas}$/$\pi r^{2}$), where $r$ is the size of a galaxy.
To estimate the total gas mass including the atomic hydrogen and molecular gas, we used the following two variations of the Kennicutt-Schmidt law, the original version \citep{Kennicutt1998}, 
\begin{equation}
\Sigma_{\rm SFR}=2.5\times10^{-4} \left( \frac{\Sigma_{{\rm M}_{\rm gas}}}{1{\rm M}_{\odot}{\rm pc}^{-1}} \right)^{1.4} {\rm M}_{\odot} {\rm yr}^{-1} {\rm kpc}^{-2},
\end{equation}
and the extended Schmidt law \citep{Shi2018},
\begin{equation}
\Sigma_{\rm SFR}=10^{-4.76}(\frac{\Sigma_{{\rm M}_{*}}^{0.5}}{1{\rm M}_{\odot}{\rm pc}^{-1}}\frac{\Sigma_{{\rm M}_{\rm gas}}}{1{\rm M}_{\odot}{\rm pc}^{-1}})^{1.09} {\rm M}_{\odot} {\rm yr}^{-1} {\rm kpc}^{-2}.
\end{equation}
The different assumptions yield slightly different gas masses for the clusters in our sample, since the extended Schmidt law has an additional $\Sigma_{M_{*}}$ term.
In both cases, the blue cluster shows the highest gas mass fraction, i.e., $f_{\rm gas}$=$M_{\rm gas}$/$M_{*}$ (Fig. \ref{newfig12}b and c), among the SDSS clusters. 
The presence of a high abundance of unusually gas-rich galaxies might explain the high blue fraction in the dynamically relaxed cluster.\\

\subsection{Filamentary structure around the blue cluster}

Why does the blue cluster have such an unusually large number of cold-gas-rich galaxies for such a massive halo?
In terms of galaxy simulations, galaxies in the blue cluster need to overcome at least two difficulties to keep star formation going after the halo mergers that lead to the assembly of the cluster at $z\sim 0.6$.
One is continuous gas feeding.
There must be a gas supply from outside of the blue cluster, because the time elapsed from cluster formation to $z=0.061$ is sufficient for member galaxies to deplete their molecular gas reservoirs and transform into a red population.
However, cosmological simulations indicate that the accretion rate of the gas into halos of this mass along cold filaments dramatically declines from $z=4$ to $z=0$ by $\sim$ 1-2 order of magnitude \citep{Keres2009}.
As a consequence, simulations (and semi-analytic models) do not predict the high blue fraction in the massive halo in the local Universe as shown in Fig. \ref{newfig9} and \ref{newfig10}.

Another key is the shock heating of the infalling gas, which is conventionally called \lq \lq hot mode accretion\rq \rq.
Even though there is a gas inflowing into the halo, the strong gravity of massive halo causes a supersonic velocity of infalling gas, which results in virial shocks \citep{Birnboim2003}.
A larger fraction of the infalling material is shock heated and reaches temperatures close to the virial temperature as the dark matter halo grows larger \citep{Rees1977}.
The shock-heated gas needs to cool in order to trigger the star formation.
In the central regions of the dark matter halos, a fraction of this hot gas is able to cool.
However the cooling of the hot gas is inefficient in halos that are more massive than the critical halo mass of $M_{\rm shock}$ = 10$^{11.7}$ M$_{\odot}$. 
This critical mass is almost constant from $z=5$ to 0 with a weak dependence on the metallicity of the intergalactic medium \citep{Dekel2006}.
The estimated halo mass of the blue cluster, 2.0$^{+1.9}_{-1.0} \times 10^{14}$ M$_{\odot}$, is well above this threshold value.
Therefore, it is unlikely for the hot gas accretion to contribute to the current star formation in the blue cluster through the cooling process in the centre.

A more realistic scenario might be filamentary inflows of cold gas, i.e, \lq \lq cold streams\rq \rq.
This process is less affected by the shock heating, and can penetrate into individual galaxies. 
Galaxy simulations confirmed that the cold streams directly contribute to the star formation \citep{Dekel2006,Dekel2009,Keres2009,Faucher2011}.
This scenario has been proposed as the main driver of galaxy formation in massive halos ($\geqq$10$^{11.7}$ M$_{\odot}$) in the high redshift Universe.
Indeed a large gas filament illuminated by a bright quasar at $z\sim 2.3$ has been discovered observationally \citep{Cantalupo2014}, which extends beyond the virial radius of the dark matter halo.
The geometry of the filament suggests an accretion flow \citep{Martin2015}.
Metal absorption lines found in spectra of galaxies at $0.4 < z < 1.4$ also suggest cold streams, though the observed physical scale is much smaller than the halo scale \citep{Rubin2012,Martin2012}.
Cold streams might also occur in the blue cluster.

In Fig. \ref{newfig8}a, projected filamentary structures in the galaxy distribution identified by \citet{Tempel2014} are shown with blue lines.
These filaments coincide with high density structure (shown by contours) implying the presence of Mpc-scale filamentary structure around the blue cluster.
We note that not all the SDSS clusters are at the nodes of filaments, probably because different algorithms are applied to cluster and filament identifications and these algorithms are not always perfect.
The blue fraction is also high along these filament (Fig. \ref{newfig8}b), suggesting that abundant gas exists not only in the blue cluster but also in the filament.
By assuming the Kennicutt-Schmidt law, the total amount of cold gas in galaxies included in a region of $6\times6$ Mpc$^{2}$ and $\pm$ 1000 km s$^{-1}$ around the blue cluster is estimated to be 10$^{11.0}$ M$_{\odot}$,  $\sim$25\% of which is in the blue cluster members.
The huge filament might be able to supply gas into the blue cluster, sufficient to sustain star formation activity in the blue cluster over the past $\sim$ 5.2 Gyr.
However, an analytic argument suggests that the cold streams cannot occur in massive halos in the local Universe, as follows \citep{Dekel2006,Dekel2009}. 
The cold streams need to overcome the shock heating in the massive halo in order to invoke the star formation.
In this case, the gas cooling time scale must be smaller than the compression time scale of the shock.
Taking this situation into account \citet{Dekel2006} calculated the maximum halo mass, $M_{\rm stream}(z)$, that allows cold streams to exist in a halo:
\begin{equation}
M_{\rm stream}(z)\sim \frac{M_{\rm shock}^{2}}{fM_{\rm PS}(z)},
\end{equation}
where $f$ and $M_{\rm PS}(z)$ are a constant factor of a few order and the Press-Schechter mass that corresponds to the typical halo mass forming at a given epoch respectively \citep{Press1974}.
Here $M_{\rm stream}(z)$ is a strong function of redshift due to the term $M_{\rm PS}(z)$.
At the redshift of the blue cluster, $M_{\rm stream}(z=0.061)$ is $\sim$ $10^{10}$ M$_{\odot}$.
Although $M_{\rm shock}$ weakly depends on the metallicity in the intergalactic medium \citep{Dekel2006}, the dependency can change $M_{\rm stream}(z=0.061)$ by less than 2 order of magnitude, within a reasonable metallicity variation. 
Therefore, the halo mass of the blue cluster, 2.0$^{+1.9}_{-1.0}\times$10$^{14}$ M$_{\odot}$, is higher than $M_{\rm stream}$, in which case, the cold streams do not survive.
The cold streams in massive halos ($M_{\rm halo}\geq$ $M_{\rm shock}=10^{11.7}$ M$_{\odot}$) are only possible at higher redshift \citep{Dekel2006,Dekel2009}. 
Based on this argument, the cold streams in a halo as massive as the blue cluster should have already disappeared in the simulated local Universes.

So far, no scenario we have considered provides us with a reasonable explanation for such a high blue fraction in a cluster in local Universe.
The calculated space density of the blue cluster is extremely low, i.e., $1/V_{\rm survey}$=$2.6 \times {10^{-8}}$ Mpc$^{3}$, where $V_{\rm survey}$ is the survey volume of our sample.
The blue cluster is probably an extreme physical case, unexpected in current theoretical frameworks of galaxy formation.
Although it is rare, future semi-analytic models and simulations need to be able to explain the presence of such a blue cluster. \\

\subsection{Multi-wavelength data}

Multi-wavelength data from X-ray to radio are helpful to quantify how special the blue cluster is in several different aspects. 
We found no bright X-ray detection in ROSAT all sky survey catalogue \citep{Voges1999,Boller2016} within the virial radius of the blue cluster.
\citet{Wang2014} measured ROSAT X-ray fluxes at positions of SDSS DR7 galaxy clusters down to S/N $\sim$ 1 level. 
In their catalogue we found an X-ray flux measurement at the position of the blue cluster, although the S/N is only 1.27 \citep{Wang2014}.
The X-ray position is offset from the centre of the blue cluster by 2.85 arcmin, which is within the virial radius of 8.7 arcmin on sky.
The observed X-ray flux of $0.663 \times 10^{-15}$ W m$^{-2}$ corresponds to the luminosity of $\log L_{X}$ = 42.8 erg s$^{-1}$.
The scaling relation between halo mass and X-ray luminosity \citep{Wang2014} provides an expected X-ray luminosity of the blue cluster.
The expected X-ray luminosity is $\log L_{X}$=43.1$^{+0.5}_{-0.5}$ erg s$^{-1}$ including the halo mass uncertainty.
Therefore the X-ray luminosity of the blue cluster is not in conflict with the scaling relation found for other local clusters.
However, it is possible that the actual X-ray luminosity is fainter than this estimate, because the SN of X-ray flux is poor.
If cold streams dominate the intercluster medium of the blue cluster, the X-ray luminosity might be much fainter, because X-ray emitting hot gas is less abundant.

Sunyaev-Zel'dovich (SZ) effect \citep{Zeldovich1969,Sunyaev1970} is another approach to detect hot plasma in the intercluster medium.
We checked the Planck SZ catalogue \citep{Planck2016} around the blue cluster and found no counterpart.
This is probably because the halo mass of the blue cluster almost approaches the detection limit of Planck SZ clusters, i.e., 20\% completeness for $\sim 10^{14}$ M$_{\odot}$ clusters at $z \sim 0.05$ \citep{Planck2016}.
The observationally suggested cold streams in the blue cluster might also weaken the SZ effect as suggested in the case of X-ray.

The current multi-wavelength data of X-ray and SZ effect are not deep enough to give a conclusive answer.
Future follow up observations are necessary to reveal special physical properties of the blue cluster from the point of view of multi-wavelength data.

\subsection{Analogy to high redshift clusters}

The physical origin of the blue cluster is a mystery.
However, it presents a useful case study as an analog of high redshift clusters.
If we simply assume that the SFR$_{\rm total}$ of the blue cluster is balanced with the gas supply, the gas accretion rate that contributes to star formation is 24.9 M$_{\odot}$ yr$^{-1}$.
The simulations predict that the accretion rate is roughly proportional to the halo mass at the same redshift \citep{Correa2018}.
Therefore, the SFR$_{\rm total}$ divided by the halo mass is an indicator of how much gas has been accreted recently by galaxies in the halo.
Observationally, SFR$_{\rm total}$/$M_{\rm halo}$ in $\sim10^{14}$M$_{\odot}$ clearly declines from $z=1$ to 0 \citep{Bai2009,Koyama2010,Popesso2012,Webb2013,Haines2013}.
It is also empirically suggested that SFR$_{\rm total}$/$M_{\rm halo}$ depends not only on the redshift but also $M_{\rm halo}$ \citep{Bai2009,Koyama2010}.

\begin{figure}
	\includegraphics[width=\columnwidth]{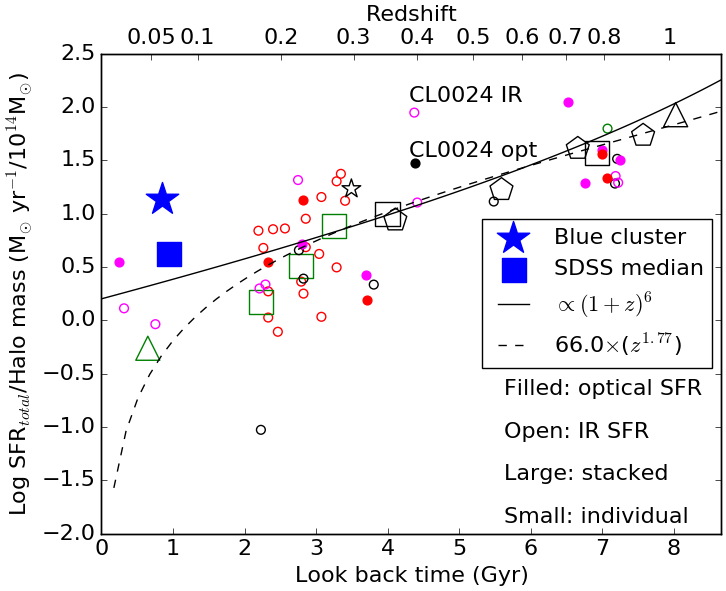}
    \caption{
    The total SFR in individual clusters divided by the halo mass as a function of the look back time or redshift.
    Only clusters hosted by massive halos ($M_{\rm halo} \geq 10^{14}$ M$_{\odot}$) are included.
    The blue cluster is shown with the blue star.
    Other data are collected from the individual cluster observations (small marks) and the stacked data (large marks). 
    The filled marks indicate the SFR derived from the optical observations, while the open marks means the infrared-derived SFR.
    The solid line is the empirical trend for the optical SFR$_{\rm total}$/$M_{\rm halo}$ \citep{Koyama2010}.
    The dashed line is the best-fit function for the infrared SFR$_{\rm total}$/$M_{\rm halo}$ \citep{Popesso2012}.
    The details of the data points are as follows.
    (Open small red circle) individual IR SFR by \citet{Haines2013}, (Open large green square) stacked IR SFR by \citet{Haines2013}, (Open large green triangle) stacked IR SFR of local clusters by \citet{Haines2013}, (Open small black circle) individual IR SFR by \citet{Popesso2012}, (Open large black square) IR SFR stacked for clusters in the COSMOS field by \citet{Popesso2012}, (Open large black triangle) IR SFR stacked for clusters in the GOODS field by \citet{Popesso2012}, (Open small black star) IR SFR of the \lq\lq Bullet cluster\rq\rq\ by \citet{Popesso2012}, (Open small magenta circle) individual IR SFR by \citet{Bai2009}, (Open small green circle) individual IR SFR by \citet{Koyama2010}, (Open large black pentagon) stacked IR SFR by \citet{Webb2013}, (Filled small magenta dot) individual optical SFR by \citet{Bai2009}, (Filled small red dot) individual optical SFR by \citet{Koyama2010}, and (Filled small black dot) individual optical SFR by \citet{Kodama2004}.
    }
    \label{newfig13}
\end{figure}

To factor out this dependence on halo mass, we compared the SFR$_{\rm total}$/$M_{\rm halo}$ of the blue cluster with other clusters hosted by  massive halos, i.e., $M_{\rm halo} \geq 10^{14}$ M$_{\odot}$ (Fig. \ref{newfig13}).
Note that Fig. \ref{newfig13} includes SFRs derived from the optical and infrared observations with different selection criteria, 
while the different initial mass functions assumed in the literature are all scaled to that of \citet{Kroupa2002}.
Infrared SFRs by \citet{Popesso2012} and \citet{Haines2013} (open large green square/triangle and others in Fig. \ref{newfig13}; see caption for details) only include the total SFRs contained within cluster luminous infrared galaxies (LIRGs: $L_{\rm IR}>10^{11}$ L$_{\odot}$, or SFR $>$ 10 M$_{\odot}$ yr$^{-1}$).
The range of star formation rates at a fixed stellar mass is not as wide as that in our analysis and other comparison samples in Fig. \ref{newfig13}.
At the lowest redshifts ($z<0.2$) very few star-forming galaxies are LIRGs.
They contribute only a small fraction of the overall SFR among star-forming galaxies, either in clusters or the field \citep[e.g.,][]{Haines2013}. 
Hence the total SFRs in LIRGs will be much lower than the total SFRs in the wider star-forming galaxy population.

Even though there could be systematic differences between the optical and infrared SFRs and large uncertainty in halo mass estimates, they broadly follow the same declining trend from high $z$ to the present. 
A deviation of the empirical trend for the optical SFR$_{\rm total}$/$M_{\rm halo}$ (solid line in Fig. \ref{newfig13}) from that in the infrared (dashed line in Fig. \ref{newfig13}) becomes larger at $z\leq$ 0.1.
This might be due to the different methods used to derive the SFR as mentioned above or the 
small-number statistics of the stacked data point at $z=0.05$ for the infrared sample \citep{Haines2013} (large green open triangle in Fig. \ref{newfig13}).
The infrared stacked data at $z=0.05$ might be biased to  extremely massive and thus evolved systems, because \citet{Haines2013} stacked 7 local clusters of which 6 are members of the Coma and Shapley super clusters. 
The remaining cluster, A3266, is also hosted by a very massive halo with 1.5$\times$10$^{15}$ M$_{\odot}$ \citep{Ettori2019}.

Apart from the global trend of SFR$_{\rm total}$/$M_{\rm halo}$, two clusters have exceptionally high accretion rates, i.e., the blue cluster and a cluster at $z=0.4$ (CL0024) \citep{Kodama2004}.
Because of this similarity, we briefly compare the blue cluster to CL0024.
The halo mass of the CL0024 cluster is $\sim$ 10$^{15}$ M$_{\odot}$ \citep{Kneib2003,Umetsu2010}, one order of magnitude larger than the blue cluster.
The expected formation epoch of the CL0024 halo in $\Lambda CDM$ is $z\sim0.5$ \citep{Ishiyama2015}. 
This is only $\sim$ 1 Gyr before the time corresponding to the cluster as observed.
In fact, the X-ray and dynamical data suggests that the cluster is in a post-collision state just 2-3 Gyr after the merging that led to its formation \citep{Umetsu2010}.
The dynamical state of CL0024 therefore seems to be different from that of the blue cluster.
It is possible that the star formation in CL0024 is triggered by the cluster-cluster major merger, rather than the cold streams in the dynamically relaxed system as observationally suggested in the blue cluster.

Among clusters with $M_{\rm halo} \geq 10^{14}$ M$_{\odot}$ at $z \leq 0.1$, the blue cluster shows the highest accretion rate, which corresponds to the typical value at higher redshift of $z\sim0.4-0.5$.
Therefore, the blue cluster could be an archetype of clusters at $z\sim0.4-0.5$, an epoch of transition in the overall galaxy population between active star formation and present-day quiescence.
Thanks to the proximity of the blue cluster, more detailed studies are possible than those at higher redshifts, e.g., X-ray observations could directly reveal the dynamical state of the cluster.
By further investigating this cluster, we will have an opportunity to understand physics we have not been aware of.

\section{Conclusions}
\label{conclusion}
We discovered the \lq \lq blue\rq \rq\ galaxy cluster hosted by a massive halo with $2.0^{+1.9}_{-1.0} \times 10^{14}$ M$_{\odot}$ at $z=0.061$.
This blue cluster is composed of 12 blue star-forming and 9 red quiescent galaxies selected from the SDSS DR7.
The X-ray flux measured at the position of the blue cluster is not in conflict with a scaling relation between X-ray luminosity and halo mass, although the X-ray detection is marginal, i.e., S/N=1.27.
The blue fraction is 0.57, placing the cluster 4.0 $\sigma$ beyond the SDSS comparison clusters under the same selection criteria. 
The blue fraction has a significance of more than 4.7 $\sigma$ when compared with semi-analytic galaxy evolution models.
The probability to find such a high blue fraction in an individual cluster is only 0.003\%, which poses a great challenge to the current standard frameworks of the galaxy formation and evolution in the $\Lambda$CDM Universe.
The galaxy distribution around the blue cluster suggests the existence of the filamentary cold gas streams in a massive halo even in the local Universe, which have already disappeared in the theoretically simulated local Universes.

\section*{Acknowledgements}
We are very grateful to the referee for many insightful comments.
We thank Hironao Miyatake for helpful comments on SZ counterparts of optically identified clusters.
We are grateful to Ho Seong Hwang for valuable discussions on the X-ray luminosity of the blue cluster.
TG acknowledges the supports by the Ministry of Science and Technology of Taiwan through grants 105-2112-M-007-003-MY3 and 108-2628-M-007-004-MY3.
We are grateful to Andrew Cooper for his assistance in proof-reading the paper.




\bibliographystyle{mnras}
\bibliography{BC_MNRAS} 



\appendix
\section{Halo mass conversion of $\nu^{2}$GC}
\begin{figure}
	\includegraphics[width=\columnwidth]{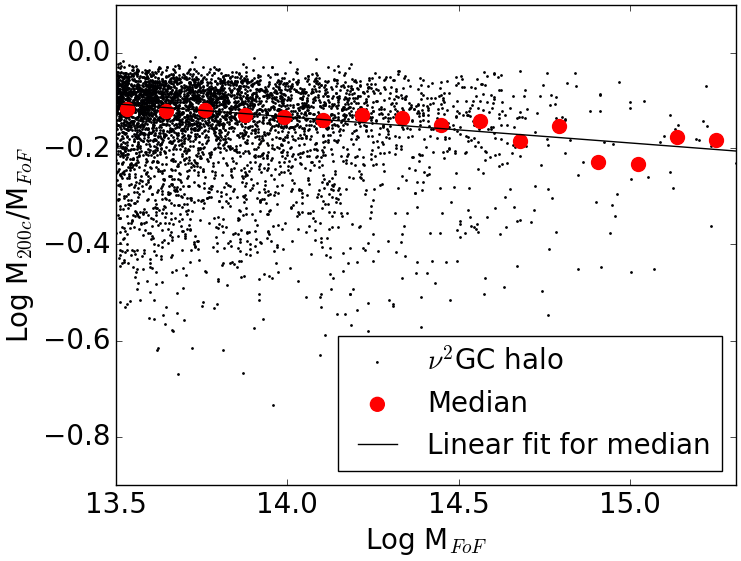}
    \caption{
    $M_{\rm 200c}$/$M_{\rm FoF}$ as a function of $M_{\rm FoF}$ in $\nu^{2}$GC dark matter simulations \citep{Ishiyama2015}.
    $M_{\rm 200c}$ is a halo mass enclosed by a radius, $R_{\rm 200c}$, where $R_{\rm 200c}$ refers to the radius at which the averaged cluster mass density is 200 times the critical density of the Universe.
    $M_{\rm FoF}$ is a total mass of dark matter particles identified by the friends-of-friends method \citep{Ishiyama2015}.
    Black dots are dark matter halo data.
    Medians of the data distribution is demonstrated by red dots.
    The medians are fitted by a linear function (solid line).
    }
    \label{newfigA1}
\end{figure}

Here we estimate a conversion formula from $M_{\rm FoF}$ to $M_{\rm 200c}$ of the $\nu^{2}$GC dark matter simulation \citep{Ishiyama2015}. 
We selected 11,394 dark matter halos with $M_{\rm FoF} > 10^{13.5}$ M$_{\odot}$ from $\nu^{2}$GC. 
$M_{\rm 200c}$ was calculated for each halo. 
Fig. \ref{newfigA1} shows log $M_{\rm 200c}$/$M_{\rm FoF}$ as a function of log $M_{\rm FoF}$.
A linear function was fitted for the median distribution of the data. 
The best fit function is
\begin{equation}
\log M_{\rm 200c}/M_{\rm FoF} = 0.628-0.054 \times \log M_{\rm FoF}.
\end{equation}
Following the formula, we converted $M_{\rm FoF}$ to $M_{\rm 200c}$ to obtain $M_{\rm halo,true}$ in Fig \ref{newfig6}. 

\section{Properties of member galaxies of the blue cluster}
Here we show physical properties, optical images, and spectra of the member galaxies in the blue cluster.

\begin{table*}
	\centering
	\caption{
	Physical parameters of member galaxies of the blue cluster.
    }
	\label{tab1}
	\begin{flushleft}
	\begin{tabular}{|l|c|c|c|c|c|c|c|c|c|}\hline
(1)&	(2)     &  (3)                 & (4)          & (5)         & (6)       & (7)      & (8)             & (9)             & (10) \\ \hline
ID&ObjId       &  SpecObjId           &  RA          &  Dec        &  Redshift & Mr       & logSFR          &  log$M_{*}$     & SF/QS \\ 
    &        &                      & (deg)        &  (deg)      &           & (mag)      & (M$_{\odot}$ yr$^{-1}$) & (M$_{\odot}$)&       \\ \hline
1& 588010136266277051  &  155878996997832704  &  136.810196  &  52.062061  &  0.05952  &  -21.00  &  0.19$\pm$0.43  &  10.57$\pm$0.10  &  SF \\
2& 588010136266342547  &  155597493625683968  &  136.953033  &  52.095886  &  0.06053  &  -21.51  &  0.15$\pm$0.47  &  10.97$\pm$0.10  &  SF \\
3& 588010136266342551  &  155878996813283328  &  136.904205  &  52.142735  &  0.06044  &  -20.97  &  0.40$\pm$0.20  &  10.40$\pm$0.09  &  SF \\
4& 588010136266277052  &  155597493613101056  &  136.810883  &  52.064072  &  0.06025  &  -20.34  &  0.14$\pm$0.50  &  10.34$\pm$0.10  &  SF \\
5& 588010136266342537  &  155597493571158016  &  136.839966  &  52.167648  &  0.06105  &  -20.49  &  0.20$\pm$0.28  &  9.86$\pm$0.07  &  SF \\
6& 588009366935568514  &  155878997295628288  &  136.911804  &  51.866158  &  0.06301  &  -21.01  &  0.64$\pm$0.15  &  10.44$\pm$0.09  &  SF \\
7& 588009366935568516  &  155878997253685248  &  136.934921  &  51.853630  &  0.06173  &  -20.49  &  0.18$\pm$0.16  &  10.18$\pm$0.10  &  SF \\
8& 588009366935634146  &  155597493604712448  &  137.081085  &  52.033100  &  0.05984  &  -20.27  &  -0.40$\pm$0.47  &  9.86$\pm$0.10  &  SF \\
9& 588010136266342441  &  155878996985249792  &  136.783630  &  52.168465  &  0.06240  &  -21.58  &  0.62$\pm$0.27  &  10.39$\pm$0.08  &  SF \\
10& 588009366935634159  &  155878996838449152  &  137.009247  &  52.118572  &  0.06273  &  -21.11  &  0.20$\pm$0.19  &  10.51$\pm$0.09  &  SF \\
11& 588010136266211504  &  155878997455011840  &  136.625397  &  51.991421  &  0.06348  &  -21.65  &  0.41$\pm$0.32  &  10.69$\pm$0.10  &  SF \\
12& 588010136266211505  &  155878997434040320  &  136.676254  &  51.958103  &  0.06286  &  -20.59  &  -0.04$\pm$0.53  &  10.73$\pm$0.11  &  SF \\
13& 588009366935568497  &  155597493608906752  &  136.915878  &  51.852680  &  0.06255  &  -21.51  &  -0.86$\pm$0.67  &  10.77$\pm$0.08  &  QS \\
14& 588010136266276968  &  155878997442428928  &  136.654922  &  51.985653  &  0.06160  &  -21.88  &  -0.96$\pm$0.75  &  10.91$\pm$0.09  &  QS \\
15& 588010136266342446  &  155878996867809280  &  136.906723  &  52.115166  &  0.05987  &  -22.32  &  -0.92$\pm$0.76  &  11.11$\pm$0.09  &  QS \\
16& 588009366935634107  &  155878996800700416  &  136.914688  &  52.067802  &  0.05752  &  -20.91  &  -1.40$\pm$0.70  &  10.51$\pm$0.09  &  QS \\
17& 588009366935634138  &  155878996809089024  &  136.979553  &  52.082661  &  0.06258  &  -20.78  &  -1.19$\pm$0.68  &  10.78$\pm$0.08  &  QS \\
18& 588009366935634114  &  155878996817477632  &  136.945343  &  52.071552  &  0.06236  &  -22.70  &  -0.76$\pm$0.73  &  11.36$\pm$0.08  &  QS \\
19& 588009366935633929  &  155878997299822592  &  136.857300  &  52.019890  &  0.06024  &  -21.03  &  -1.36$\pm$0.69  &  10.52$\pm$0.09  &  QS \\
20& 588010136266342573  &  155597493587935232  &  136.920883  &  52.158199  &  0.05736  &  -20.40  &  -1.08$\pm$0.65  &  10.13$\pm$0.11  &  QS \\
21& 588010136266342576  &  155878996834254848  &  136.934174  &  52.154041  &  0.06128  &  -20.69  &  -0.58$\pm$0.46  &  10.30$\pm$0.10  &  QS \\ \hline
    \end{tabular}\\
    Column (1) ID assigned in this paper. (2) SDSS DR7 imaging ID. (3) SDSS DR7 spectroscopic ID. (4) and (5) Right ascension and declination of SDSS DR7 coordinates. (6) Spectroscopic redshift. (7) Absolute magnitude in $r$ band. (8) Star-formation rate \citep{Brinchmann2004}. (9) Stellar mass \citep{Kauffmann2003}. (10) Separation between blue star-forming (SF) and red quiescent galaxies (QS) based on Fig. \ref{newfig4}. 
    \end{flushleft}
\end{table*}


\begin{figure*}
	\includegraphics[width=5.0in]{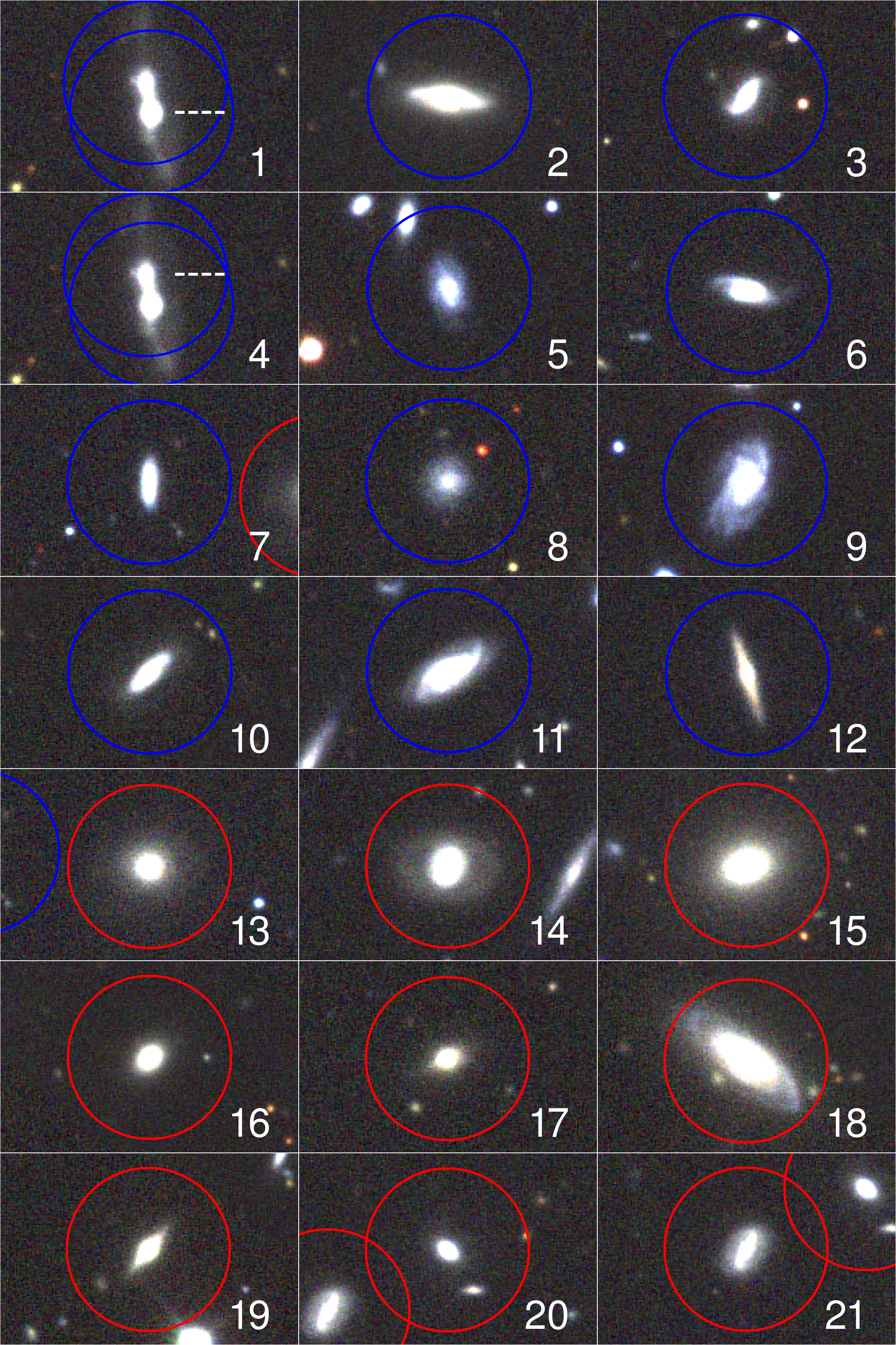}
    \caption{
    Member galaxies of the blue cluster.
    Blue star-forming and red quiescent galaxies are marked by blue and red circles, respectively.
    Member galaxies 1 and 4 are marked by white dashed lines.
    RGB colour was created from $r$, $i$, and $g$ bands of SDSS DR7 images.
    }
    \label{newfigB1}
\end{figure*}

\begin{figure*}
	\includegraphics[width=6.0in]{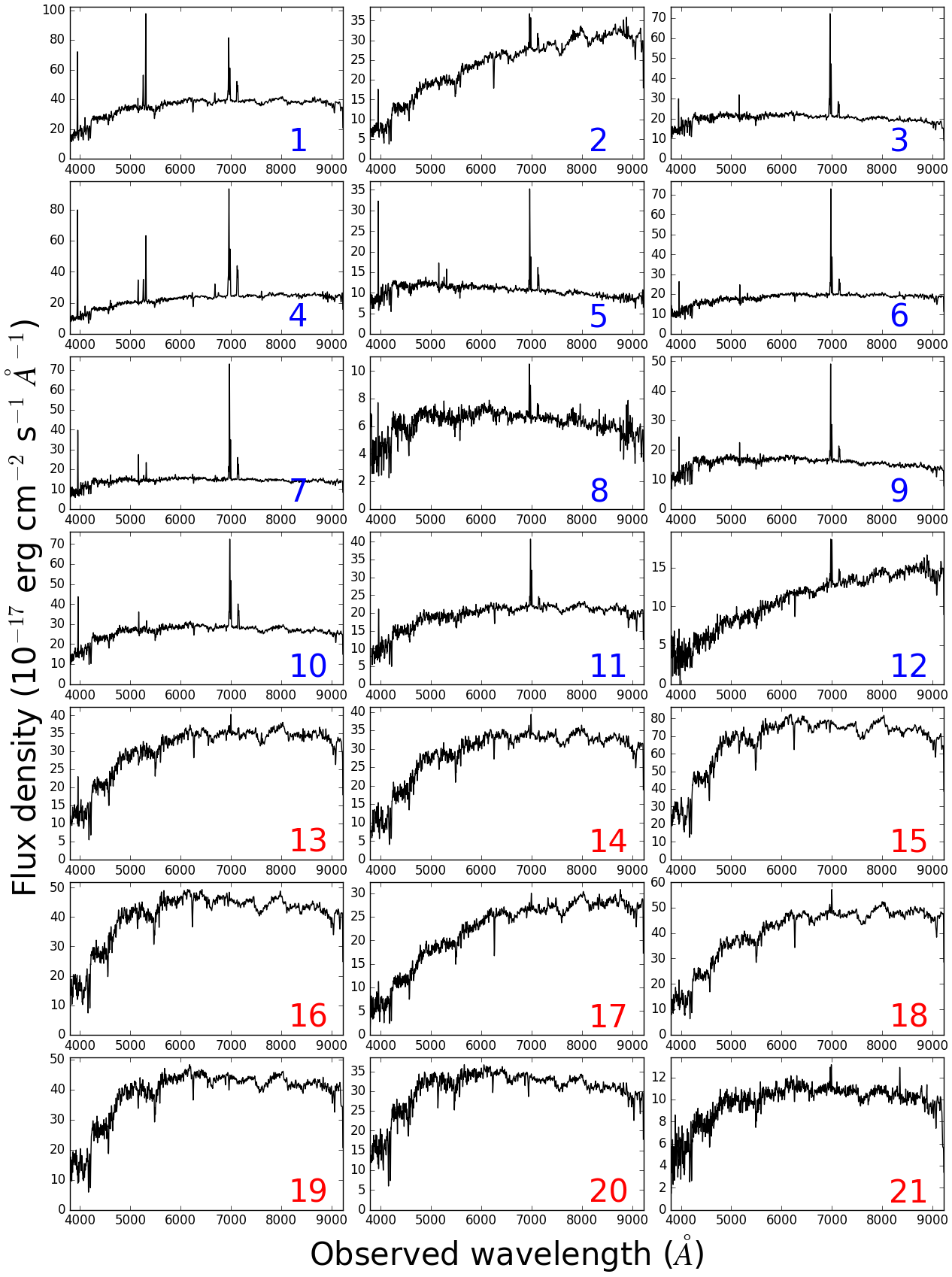}
    \caption{
    Spectra of member galaxies of the blue cluster.
    Blue star-forming and red quiescent galaxies are labeled by blue and red IDs, respectively.
    }
    \label{newfigB2}
\end{figure*}

\bsp	
\label{lastpage}
\end{document}